\documentclass{article}

\usepackage{arxiv}

\usepackage[utf8]{inputenc} % allow utf-8 input
\usepackage[T1]{fontenc}    % use 8-bit T1 fonts
\usepackage{hyperref}       % hyperlinks
\usepackage{url}            % simple URL typesetting
\usepackage{booktabs}       % professional-quality tables
\usepackage{amsfonts}       % blackboard math symbols
\usepackage{nicefrac}       % compact symbols for 1/2, etc.
\usepackage{microtype}      % microtypography
\usepackage{lipsum}		% Can be removed after putting your text content
\usepackage{graphicx}
\usepackage{doi}

\hypersetup{
pdftitle={Crosslinking degree variations enable programming and controlling soft fracture via sideways cracking},
pdfsubject={cond-mat.mtrl-sci},
pdfauthor={Miguel Angel Moreno-Mateos, Paul Steinmann},
pdfkeywords={Soft fracture, Elastomers, Phase-field models, Mechanics of materials},
}

\hypersetup{
    colorlinks=true,
    linkcolor=blue,     % Color for internal links
    filecolor=magenta,  % Color for file links
    urlcolor=black,      % Color for URLs
    citecolor=blue,    % Color for citations
}

\usepackage{booktabs}
\usepackage{amssymb}
\usepackage{amsmath}
\usepackage{stmaryrd}
\usepackage{mathrsfs}
\usepackage{color}
\usepackage{graphicx}
\usepackage{latexsym}
\usepackage{tabls}
\usepackage{graphicx}
\usepackage{epsfig}
\usepackage{subfigure}
\usepackage{rotating}
\usepackage{latexsym}
\usepackage[mathcal]{eucal}
\usepackage{amssymb}
\usepackage{colortbl}
\usepackage{times}
\usepackage{longtable}
\usepackage{algorithmic}
\usepackage{algorithm}
\usepackage{psfrag}
\usepackage{tabularx}
\usepackage{epstopdf}
\usepackage{gensymb}
\usepackage{bbm}
\usepackage{setspace}
\usepackage{fullpage}
\usepackage{url}
\usepackage{calc}
\usepackage{siunitx}

\usepackage{lineno,hyperref}
\usepackage{stmaryrd}
\usepackage{subfigure}
\modulolinenumbers[50]
\usepackage{epstopdf}
\usepackage{array}
\usepackage{textcomp}

\usepackage{booktabs}
\usepackage{mathrsfs}
\usepackage{multirow}
\usepackage{lscape}
\usepackage{amsmath}

\usepackage{mathtools}
\usepackage{amssymb}
\usepackage{color,soul}
\usepackage{bm}

\usepackage{titlesec}
\titleformat{\paragraph}
{\normalfont\normalsize\bfseries}{\theparagraph}{1em}{}
\titlespacing*{\paragraph}
{0pt}{3.25ex plus 1ex minus .2ex}{1.5ex plus .2ex}

\usepackage{tikz}

\usepackage{makecell}
\usepackage{bbm}
\usepackage{float}

\usepackage{lineno}

\usepackage[font=scriptsize]{caption}
\def\vc #1{\mbox{\boldmath $#1$}}

\begin{document}

\title{Crosslinking degree variations enable programming and controlling soft fracture via sideways cracking}

\author{ \href{https://orcid.org/0000-0002-3476-2180}{\includegraphics[scale=0.06]{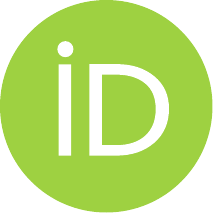}\hspace{1mm}Miguel Angel Moreno-Mateos}\thanks{Corresponding author.} \\
	Institute of Applied Mechanics\\
	Friedrich-Alexander-Universität Erlangen–Nürnberg\\
	91058, Erlangen, Germany\\
	\texttt{miguel.moreno@fau.de} \\
	%% examples of more authors
	\And
	\href{https://orcid.org/0000-0003-1490-947X}{\includegraphics[scale=0.06]{orcid.pdf}\hspace{1mm}Paul Steinmann} \\
	Institute of Applied Mechanics\\
	Friedrich-Alexander-Universität Erlangen–Nürnberg\\
	91058, Erlangen, Germany\\
	Glasgow Computational Engineering Centre\\
	University of Glasgow\\
	G12 8QQ, United Kingdom\\
}

\maketitle

\begin{abstract}
Large deformations of soft materials are customarily associated with strong constitutive and geometrical nonlinearities that originate new modes of fracture. Some isotropic materials can develop strong fracture anisotropy, which manifests as modifications of the crack path. Sideways cracking occurs when the crack deviates to propagate in the loading direction, rather than perpendicular to it. This fracture mode results from higher resistance to propagation perpendicular to the principal stretch direction. It has been demonstrated that such fracture anisotropy is related to the microstructural stretch of the polymer chains, also known as strain crystallization mechanisms. However, the precise variation of the fracture behavior with the degree of crosslinking is not fully understood. Leveraging experiments and computational simulations, here we show that the tendency of a crack to propagate sideways in the two component Elastosil P7670 increases with the degree of crosslinking. We explore the mixing ratio for the synthesis of the elastomer that establishes the transition from forward to sideways fracturing. To assist the investigations, we construct a novel phase-field model for fracture where the critical energy release rate is directly related to the crosslinking degree. Our results demonstrate that fracture anisotropy can be programmed during the synthesis of the polymer. Then, we propose a roadmap with composite soft structures with low- and high-crosslinked phases that allow for control over fracture,  arresting and/or directing the fracture. By extending our computational framework as a virtual testbed, we capture the fracture performance of the composite samples and enable predictions based on more intricate composite unit cells. Overall, our work offers promising avenues for enhancing the fracture toughness of soft polymers.
\end{abstract}


\section{Introduction}
Soft materials form the basis of disruptive applications in soft robotics, bioengineering, microfluidics, and flexible electronics, among others. In the catalog of soft materials, elastomers offer responses at large deformations with stiffness (Young's) modulus low as only a few kilopascal \cite{Moreno2021}.  The constitutive response of elastomers is influenced by their microstructural features, particularly the polymer chains and their crosslinked networks \cite{Ghareeb2020}. Significant research has focused on designing the microstructure of these networks to optimize their effective response \cite{Cai2022}. In the rapidly evolving field of elastomeric materials, understanding their fracture mechanics is crucial. Elastomers display a wide array of failure modes \cite{Eirich1973} and a variety of unstable cracks has been documented, including branching cracks and rough-surface cracks \cite{Bowden1967}, oscillatory cracks \cite{Chen2017i}, shark-fin-like cracks \cite{Hamm2020}, spiral cracks \cite{Leung2001}, crescent cracks \cite{Marthelot2014}, tongue-like cracks \cite{Hamm2008}, zigzag cracks \cite{Takei2013}, en-passant cracks \cite{Schwaab2018}, helical cracks \cite{Pons2010}, and static sideways cracks \cite{Gent2003, Lee2019c}, observed in natural rubber \cite{Ju2023} and collagenous tissues \cite{Toaquiza2022}, as well as fatigue-induced sideways cracks \cite{Xiang2020}. The onset of edge cracks has also been explored \cite{Xue2024} and the fatigue behavior of elastomeric materials has garnered considerable interest, particularly for their role in adhesion-based applications within bioengineering \cite{Liu2020k}. Additionally, multiphysics fracture has been investigated, including studies on magneto-active elastomers \cite{Moreno-Mateos2023} and dielectric elastomers \cite{Moreno-Mateos2024a}. Lastly, anisotropic fracture in elastomers is typically associated with materials that are inherently anisotropic in their undeformed state \cite{Gao1988}, with biological materials serving as prime examples \cite{Bircher2019, Taylor2012}.

The control of fracture properties in elastomers mainly focuses on improving damage tolerance, even achieving self-healing capabilities. The strategies to accomplish this are numerous. Among others, composite materials can direct fractures along interfaces in multi-phase systems \cite{Magrini2024}. Phase contrast with tough mesophase structures holds promise for creating self-healing materials \cite{Li2021}. Magneto-mechanical and electro-mechanical coupling mechanisms can delay crack propagation \cite{Moreno-Mateos2023, Moreno-Mateos2024a} and high-functionality cross-links can simultaneously enhance stiffness and fracture toughness \cite{Zhao2017, Lin2022}. Regarding the latter, strain-induced crystallization creates domains where polymer chains align with the principal stretch direction, leading to the hyperelastic strain-hardening behavior characteristic of elastomers \cite{Hartquist2023}. More intriguingly, this phenomenon results in distinctive fracture patterns, such as sideways cracking. Sideways cracks occur when the crack propagates in the loading direction rather than perpendicularly, due to the high resistance of fractures to propagate across these aligned domains. In the present work, we devise also control of the sideways-forward cracking behavior through the modification of the crosslinking degree. 

Theoretical models and computational frameworks hold great potential for informing experiments and guiding the design of novel materials with superior mechanical and fracture properties. The catalog of modeling approaches is extensive \cite{Cheng2024}, including, inter alia, the extended finite element method \cite{Moes1999}, cohesive zone models \cite{Marigo2023}, peridynamics \cite{Silling2000} and continuum-kinematics-inspired peridynamics \cite{Javili2019}, and configurational mechanics \cite{Schmitz2023,Moreno-Mateos2024b}. Among these, Finite Elements phase-field models are prominent to model fracture \cite{Miehe2010,Russ2020,Lo2022}. While the phase-field approach has been applied to anisotropic fracture, these extensions have primarily focused on materials that are anisotropic in their undeformed state \cite{Teichtmeister2017, Yin2020, Yin2021, Rezaei2022, Nagaraja2023}. Sideways cracking, however, is due to the anisotropic fracture behavior that develops with the deformation of elastomers that are isotropic in their original, undeformed state. To the best of our knowledge, only a few studies have addressed anisotropic fracture resulting from microstructural anisotropy that develops during deformation \cite{Schreiber2021}, and none have proposed formulations where fracture toughness, in the spirit of the critical energy release rate, increases with the deformation of the medium, preserving to the best the physical motivation.

The fracture response of highly crosslinked elastomers is complex and highly dependent on the microstructural anisotropy of polymer chains that develops with increasing deformation. The degree of crosslinking plays a pivotal role in the formation of crystal domains in the loading direction. As a consequence, high crosslinking states make elastomers stiffer, which may be a drawback for applications requiring compliant materials. The significant advances in soft materials and the rise of strategies to enhance their failure tolerance prompt the question of whether the smart design of the crosslinking degree at the synthesis stage may enable new pathways for creating soft structures capable of deviating and even arresting crack propagation. Furthermore, integrating the same elastomer in both its low- and high-crosslinked versions, in the spirit of a fiber-reinforced structure or even a composite with fracture-resistant inclusions, promises new avenues to program and control fracture in soft structures. To the best of the authors' knowledge, only a few works have explored sideways cracking in polymers in a comprehensive way. Lake et al. \cite{Lake1991} and Gent et al. \cite{Gent2003} reported the first observations of sideways fractures. Lee et al. \cite{Lee2019c} revisited sideways cracking, describing the propensity of sideways cracks to develop under different deformation rates and thicknesses of rectangular samples. Li et al. \cite{Li2020a} characterized fiber-reinforced elastomers and Cox et al. \cite{Cox2019} explored cohesive fracture at the interface in multiphase elastomer structures. Although these works provide fundamental insights into the mechanisms driving such fracture patterns, they fall short of describing their formation in relation to the crosslinking ratio of the elastomer, which can be easily adjusted through the mixing ratio of the raw phases. Moreover, no work has exploited sideways cracking to deviate and eventually arrest fractures in composite structures combining different crosslinking ratios. This knowledge gap holds significant potential for innovative solutions in fracture management of soft materials.

In this work, we aim to fill the current knowledge gap with a comprehensive experimental and computational study of sideways fracture and its potential application in designing soft, yet ultra-tough, fracture-resistant structures. To achieve this, we manufacture samples using the soft elastomer Elastosil P7670 with varying mixing ratios of the raw phases. Our results demonstrate that increasing the degree of crosslinking promotes sideways cracking, achieving maximum stiffness at the optimal mixing ratio. We also find that slow deformation rates enhance the occurrence of sideways cracking. Since the experiments alone are insufficient to describe the constitutive stress state near the crack tip, we develop a theoretical model and implement it within a Finite Element framework. This framework successfully captures the sideways fracture patterns and relates them to the material's fracture properties across all crosslinking ratios. Eventually, it provides experimental insights into the anisotropic toughening mechanisms that arise with deformation. Subsequently, we extend the model as a computational testbed to design soft composite elastomers capable of guiding and arresting fractures. The framework is based on a phase-field model, enabling us to predict the fracture resistance of composite structures. Through this integrated approach, we pave the way for innovative designs in soft materials that combine high toughness with controlled fracture behavior.

\section{Results}
\subsection{Adjusting the crosslinking degree via mixing ratio}
\begin{figure}[h!]
\centering
\includegraphics[width=1\textwidth]{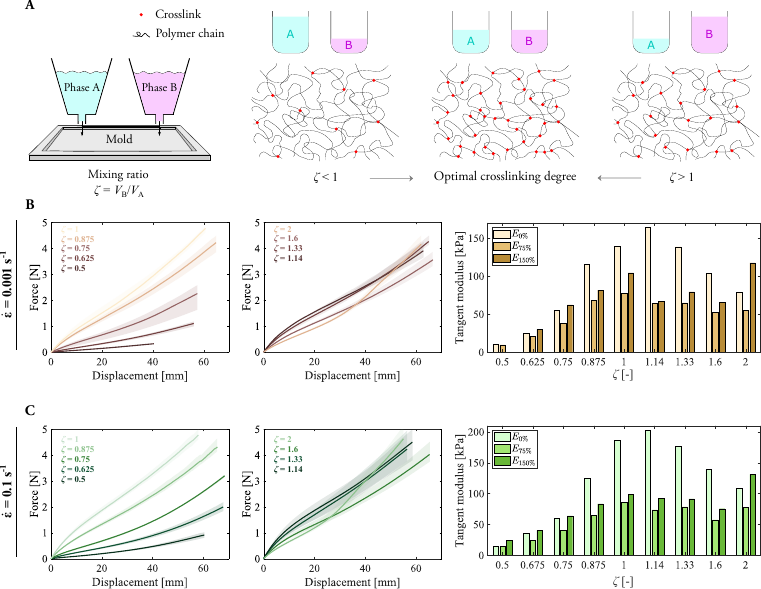}
\caption{\textbf{Experimental characterization of the mechanical behavior of uncut samples depending on the crosslinking ratio and loading rate.} Results for tensile testing on pristine (uncut) samples with mixing ratios ($\zeta$) of \qty{0.5}{}, \qty{0.625}{}, \qty{0.75}{}, \qty{0.875}{}, \qty{1}{}, \qty{1.14}{}, \qty{1.33}{}, \qty{1.6}{}, and \qty{2}{} and different loading rates. (A) Overview of the synthesis methodology of the elastomeric samples and relationship between the mixing ratio of the phases that conform the elastomer and the crosslinking degree of the polymer microstructure. For the optimal mixing ratio of 1:1, the crosslinking is maximum. (B) Results for the tensile tests on the samples with different mixing ratios at a quasi-static strain rate of $\dot{\epsilon}=\qty{0.001}{\per \second}$. (C) Results for the tensile tests on the samples at a larger strain rate of $\dot{\epsilon}=\qty{0.1}{\per \second}$. The tangent moduli at \qty{0}{\%}, \qty{75}{\%}, and \qty{150}{\%} strain are described through barplots. The shaded areas describe the scatter for three repetitions for each test condition. The maximum displacement of the curves is determined by the rupture of the samples.}
\label{fig:Exp_mixing}
\end{figure}

To investigate how fracture is modulated by the crosslinking degree, it is first necessary to describe the influence on the constitutive behavior. This description is crucial not only for calibrating material models but also for understanding how the effectiveness of crosslinking evolves with the mixing ratio of the elastomer during synthesis. The proportion between the phases, quantified in this work by the mixing volume ratio ($\zeta$), has been shown to play a fundamental role. As described in \cite{Garcia-Gonzalez2021}, among other works, there exists an optimal ratio for which the crosslinking is maximal, and below and beyond this ratio, the crosslinking degree diminishes (see Fig.~\ref{fig:Exp_mixing}.A). %Works in the literature have explored the response under different loading modes, including tensile \cite{Garcia-Gonzalez2021}, compression \cite{Cai2022}, and shear deformations \cite{Moreno-Mateos2022b}, and have substantiated applications such as self-healing based on the nature of crosslinking \cite{Li2016}.

Here, we explore the mechanical response with tensile tests for an array of mixing ratios including \qty{0.5}{}, \qty{0.625}{}, \qty{0.75}{}, \qty{0.875}{}, \qty{1}{}, \qty{1.14}{}, \qty{1.33}{}, \qty{1.6}{}, and \qty{2}{}, under quasi-static and rate-dependent load conditions. For each ratio, we calculate the tangent moduli at three engineering strains: \qty{0}{\%} ($E_{0\%}$), \qty{75}{\%} ($E_{75\%}$), and \qty{150}{\%} ($E_{150\%}$). The Young's modulus (tangent modulus $E_{0\%}$) of the elastomer deformed quasi-statically spans between \qty{10.1}{\kilo \pascal} for $\zeta=0.5$ to \qty{163.9}{\kilo \pascal} for $\zeta=1.14$, as depicted in Fig.~\ref{fig:Exp_mixing}.B. The tangent moduli $E_{75\%}=\qty{77.4}{\kilo \pascal}$ and $E_{150\%}=\qty{104.3}{\kilo \pascal}$ indicate, however, a maximum stiffness for $\zeta=1$. For $\zeta > 1$, $E_{0\%}$ and $E_{75\%}$ decrease, denoting a less efficient crosslinking. For $\zeta=2$, strong strain hardening occurs with a $E_{150\%}$ of \qty{117.4}{\kilo \pascal}. For rate-dependent deformation, as reported in Fig.~\ref{fig:Exp_mixing}.C, the Young's modulus ($E_{0\%}$) spans between \qty{14.1}{\kilo \pascal} for $\zeta=0.5$ to \qty{203.0}{\kilo \pascal} for $\zeta=1.14$. For $\zeta=1$, $E_{75\%}=\qty{85.9}{\kilo \pascal}$ and $E_{150\%}=\qty{98.7}{\kilo \pascal}$ show their maximums. Overall, the change in the stiffness with the mixing ratio, hence the crosslinking degree, is larger for $\zeta$ between 0.5 and 1 than from 1 to 2. Hereafter, we will focus on the first interval. The effect of the crosslinking degree on the constitutive behavior sets the basis for the investigation of the fracture performance in the following sections.

%%% FIGURE 1

\subsection{Tunable fracture anisotropy on isotropic elastomers}
What causes anisotropic fracture in a isotropic elastomer? The seminal investigation of Greensmith \cite{Greensmith1956} refered to anisotropic fracture in filled vulcanized rubbers knotty tearing, while Lake and co-authors reported on deviations in vulcanized rubbers, describing them as ``hammer-head'' cracks \cite{Lake1991}. Later, Gent et al. \cite{Gent2003} calculated the energy release rate for forward and sideways cracking in elastomers with large crack tip blunting and concluded that the energy release for sideways fracturing is about half the energy released in forward propagation. The authors inferred that the fracture strength at the crack tip had to be anisotropic to explain such unusual fracture patterns. More recently, Marano et al. \cite{Marano2010} and Lee et al. \cite{Lee2019c} revisited sideways fracturing and demonstrated that such anisotropy is due to the polymer network microstructure, which becomes anisotropic at large deformations. A crack perpendicular to the aligned polymer chains has to break more chains per unit length of propagation. For a same polymer, how does this change with the crosslinking ratio?

\begin{figure}[h!]
\centering
\includegraphics[width=0.89\textwidth]{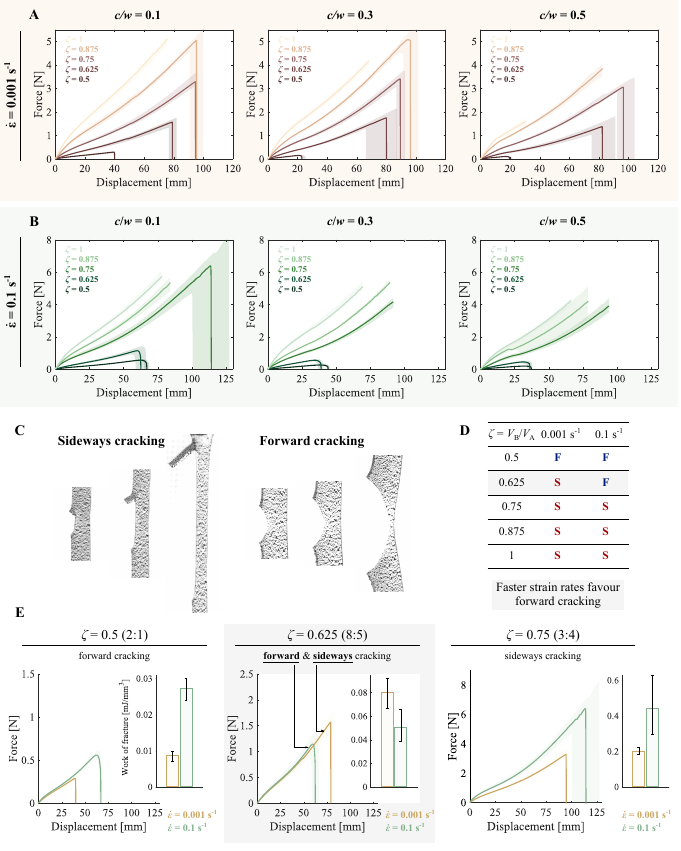}
\caption{\textbf{Experimental characterization of the fracture behavior of pre-cut samples depending on the crosslinking ratio and loading rate.} Competition of sideways cracking and forward cracking under tensile loading and depending on the mixing ratio to synthesize the elastomer and deformation rate. Samples with mixing ratios ($\zeta$) of \qty{0.5}{}, \qty{0.625}{}, \qty{0.75}{}, \qty{0.875}{}, \qty{1}{} and crack-width ratios ($c/w$) of \qty{0.1}{}, \qty{0.3}, and \qty{0.5}{} are tested. (A) Experiments at quasi-static strain rate of $\dot{\epsilon}=\qty{0.001}{\per \second}$ and (B) experiments at larger strain rate of $\dot{\epsilon}=\qty{0.1}{\per \second}$. (C) Sideways and forwards cracking on samples with $\zeta=1$ and $\zeta=0.5$, respectively, coated with a speckle pattern. (D) Overview of the fracture behavior as a function of the mixing ratio and strain rate. ``S'' denotes sideways cracking and ``F'' forward cracking. (E) Effect of the strain rate on the fracture behavior. Faster rates originate forward cracking while quasi-static deformation produces sideways cracking on a sample with $\zeta=0.625$. For small and high ratios of \qty{0.5}{} and \qty{75}{}, respectively, the loading rate does not alter the fracture pattern. The work of fracture is calculated as the area under the force--displacement curves. The shaded areas describe the scatter for three repetitions for each test condition.}
\label{fig:Exp_fracture}
\end{figure}

To understand how anisotropy in fracture strength varies with the crosslinking degree, we conduct a characterization under tensile loading of samples with initial pre-cuts and mixing ratios. We explore three different geometries featuring pre-cuts of three different lengths (see Materials and Methods~\ref{sec:materials_synthesis} and Supplementary Material (Methods 1) for the preparation of the samples). The force-displacement curves for quasi-static tests in Fig.~\ref{fig:Exp_fracture}.A indicate that forward fracturing occurs only for the mixing ratio $\zeta= \qty{0.5}{}$ and sideways fracturing for $\zeta= \qty{0.625}{}$, \qty{0.75}{}, \qty{0.875}{}, and \qty{1}{}. Illustrations of sideways and forward cracks along a propagation event are provided in Fig.~\ref{fig:Exp_fracture}.C for samples with mixing ratios of \qty{1}{} and \qty{0.5}{}, respectively. To investigate rate-dependent fracture behavior, we conduct tests with a faster strain rate of $\dot{\epsilon}=\qty{0.1}{\per\second}$. Interestingly, applying a higher strain rate to the sample with $\zeta=0.625$ produces a forward crack rather than sideways. As suggested in \cite{Lee2019c}, the potential reasons are twofold. First, viscous inelasticities at high rates hinder the separation of polymer chains, which is on the base of a sideways crack extension. Hence, the elastic energy is invested in breaking covalent bonds (cutting chains) during forward cracking. Second, high strain rates do not allow the time-dependent formation of the crystal domains that endow the elastomer with fracture anisotropy.

A summary of the type of fracture as a function of the mixing ratio and strain rate is provided in Fig.~\ref{fig:Exp_fracture}.D, and the alteration of the forward--sideways threshold towards higher mixing ratios through the deformation rate is described in Fig.~\ref{fig:Exp_fracture}.E. Here, we depict the force--displacement curves and work of fracture for the mixing ratio at which the transition occurs ($\zeta=0.625$), and the neighboring ones. While the work of fracture out of the transition mixing ratio is larger for faster deformation, for the intermediate one the formation of a forward crack at fast deformation drastically decreases it. 

Not only do we distinguish between forward and sideways propagation, but we also characterize the sideways crack's path, which varies with the mixing ratio. We provide images of the crack at the end of its sideways propagation, in both the material (undeformed) and spatial (deformed) configurations (Supplementary Material (Results 3, Figs.~S4 and S5)).

%%% FIGURE 2

\subsection{A computational continuum model captures the dependency between fracture toughness and crosslinking degree}\label{sec:comput_model_sideways}
The experimental results for fracture, whereby sideways cracking was observed, demonstrate a strong increase in fracture anisotropy with the crosslinking degree (Fig.~\ref{fig:Exp_fracture}). The rise of intricate geometrical non-linearities and the complex evolution of the microstructure under large deformations motivates the conceptualization of modeling and computational tools that replicate and inform such a behavior. In the present work, we construct a continuum phase-field model for sideways fracture at finite strains. The purpose of the model is twofold: first, to understand how fracture anisotropy develops with the deformation of the medium and crosslinking degree, and second, to provide a virtual testbed to design functional structures capable of controlling fracture.

Models in the literature have addressed fracture in anisotropic materials like fiber-reinforced or biological materials, e.g., \cite{Dorn2021f,Yin2020}. Unlike these examples, sideways cracking occurs in isotropic solids whose microstructure transitions into anisotropic arrangements with increasing deformation. Energetically-motivated fracture models propose that the strain energy in the medium competes with the energy dissipation required to create crack surfaces \cite{Francfort1998}. Here, phase-field models introduce a damage order parameter to regularize the crack discontinuity. To model anisotropic fracture, some phase-field models suggest modifying the regularized crack surface density (see $\gamma$ in Equation~S.3 in the Supplementary Material (Methods 2)), especially for strong anisotropy \cite{Taylor1998,Li2015,Schreiber2021}. 

\begin{figure}[h!]
\centering
\includegraphics[width=0.73\textwidth]{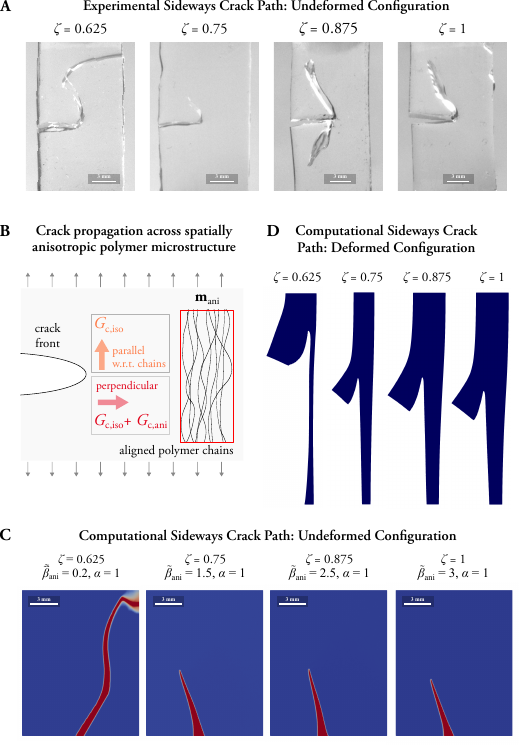}
\caption{\textbf{A computational continuum model captures the increase of the fracture anisotropy with the crosslinking degree.} The model replicates the experimentally observed sideways cracks observed from the experiments on samples with initial pre-cuts of \qty{6}{\milli \meter} ($c/w=0.5$). 
(A) Results for crack propagation for samples with mixing ratios $\zeta=0.625,0.75,0.875,1$ and initial pre-cuts in the undeformed configuration. The images correspond to full rupture for the low crosslinked sample with $\zeta=0.625$ and to maximum sideways extension for $\zeta=0.75$, \qty{0.875}{}, and \qty{1}{}. Low crosslinked samples for $\zeta=0.5$ are not included since cracks propagate forward and the path is the standard horizontal one. 
(B) Illustration of the anisotropic increase in the fracture energy release rate with the formation of anisotropic microstructural arrangements. The polymer chains and crystal domains align with the principal stretch direction, i.e., load direction and the fracture resistance increases in the perpendicular direction. 
(C) Sideways crack captured with the computational continuum model for samples with mixing ratios $\zeta=0.625,0.75,0.875,1$ in the undeformed configuration. Due to symmetry arguments only the upper half of the sample is shown. 
(D) Sideways crack captured with the computational continuum model for samples with mixing ratios $\zeta=0.625,0.75,0.875,1$ in the deformed configuration. 
Note that each crosslinking degree owes to a mixing ratio ($\zeta$) and is associated to the respective constitutive parameters described in the calibration in Supplementary Materials (Results 1).}
\label{fig:comput_frac}
\end{figure}

Unlike these existing approaches, we propose a model that preserves the purest physical notion driving sideways cracking, i.e., strain-dependent fracture anisotropy. To that end, our approach renders $\gamma$ unaltered and introduces a anisotropic critical energy release rate in the format of $G_\text{c}=G_\text{c}\left(\mathbf{F},d \right)$, with $G_\text{c}\left(\mathbf{F},d \right)=G_\text{c,iso}+G_\text{c,ani}\left(\mathbf{F},d \right)$, where $G_\text{c,ani}$ is an additional contribution activated when the material is stretched and the polymer chains align with the loading direction (see Equation~\ref{eq:gamma_crack_regularisation}). Its effect is an increase in the fracture resistance in the direction perpendicular to the load direction, as illustrated in Fig.~\ref{fig:comput_frac}.B. To mimic the increase in the fracture resistance with the deformation of the medium perpendicular to the loading direction (formation of anisotropic, crystal domains), $G_\text{c,ani}\left(\mathbf{F},d \right)$ incorporates the term $ \langle \mathbf{E}:\left[\hat{\mathbf{M}}_\text{ani}\otimes\hat{\mathbf{M}}_\text{ani}\right]\rangle_+^\alpha$, as detailed in Equation \ref{eq:beta_aniso}. The use of the Green-Lagrange strain $\mathbf{E}$ allows to amplify the anisotropic resistance with the stretch of the medium perpendicular to the loading direction $\hat{\mathbf{M}}_\text{ani}$, i.e., direction in which the chains align. Defining $G_\text{c}\left(\mathbf{F},d \right)$ as a material function rather than a constant enables us to model the evolution of anisotropy with the deformation of the medium without altering the method by which the crack is regularized. This approach preserves the physical significance of all constitutive and fracture parameters in our model. The computational modeling framework, which is introduced in Materials and Methods \ref{sec:constitutive_framework} and  thoroughly described in the Supplementary Material (Methods 2), permits to mimic the high geometrical nonlinearities, crack onset, and propagation. Thus, it allows to infer the $G_\text{c}$ function that replicates the crack paths observed from the experiments. 

Low crosslinked samples ($\zeta=0.625$) render crack path patterns that combine initial sideways propagation followed by forward extension, eventually rendering a S-shaped fracture pattern (see Fig.~\ref{fig:comput_frac}.A and Fig.~S4). Highly crosslinked samples ($\zeta=\{0.75,0.875,1\}$), however, undergo sideways extension and full rupture only occurs by brittle fracture under significant higher loading. As illustrated in Fig.~\ref{fig:comput_frac}.A and Fig.~S3, the ultra-high toughness of highly crosslinked samples ($\zeta=1$) enables long sideways crack extensions before sudden brittle fracture. For a medium crosslinked sample ($\zeta=0.75$), the sideways crack path is reduced due to the earlier brittle fracture. The S-shaped pattern that occurs in samples with $\zeta=0.625$ is modeled for a moderate anisotropy amplification factor $\tilde{\beta}_\text{ani}$ of 0.2 while the sideways pattern for a highly crosslinked sample with $\zeta=1$ requires a value of 3, i.e., 15 times larger (see Fig.~\ref{fig:comput_frac}.D). Due to the difference in the constitutive response (see results for uncut samples in Supplementary Material (Results 1)), the samples with $\zeta$ of \qty{0.75}{} and \qty{0.875}{} require intermediate amplification values of 1.5 and 2.5, respectively. For the reader interested in the dependency on $\tilde{\beta}_\text{ani}$, we report additional results in the Supplementary Material (Results 4). Overall, these findings highlight that fracture anisotropy scales with the crosslinking degree or, equivalently, with the mixing ratio up to the optimal $\zeta=1$ for maximum crosslinking.

%%% FIGURE 3

\subsection{Engineered composite elastomers allow to program \& control fracture}
Based on the above capacities to program the fracture response at the material level, we go a step further and conceptualize composite structures able to  control the crack path and increase the fracture resistance. To preserve the effective soft nature of the structure, we select a low-crosslinked phase with mixing ratio $\zeta=0.5$. To control the direction of fracture extending in the soft low-crosslinked phase, which occurs as forward fracturing as described in Fig.~\ref{fig:Exp_fracture}, a second high-crosslinked phase with $\zeta=1$ is incorporated as fiber-like domains. To do so, composite samples are fabricated with an open mold and an auxiliary 3D-printed mold to cast the fibers attending to the desired topology. Both phases polymerize together to assure strong crosslinking at the interface. The method is fully described in Materials and Methods~\ref{sec:materials_synthesis_composite}. 

\begin{figure}[h!]
\centering
\includegraphics[width=1\textwidth]{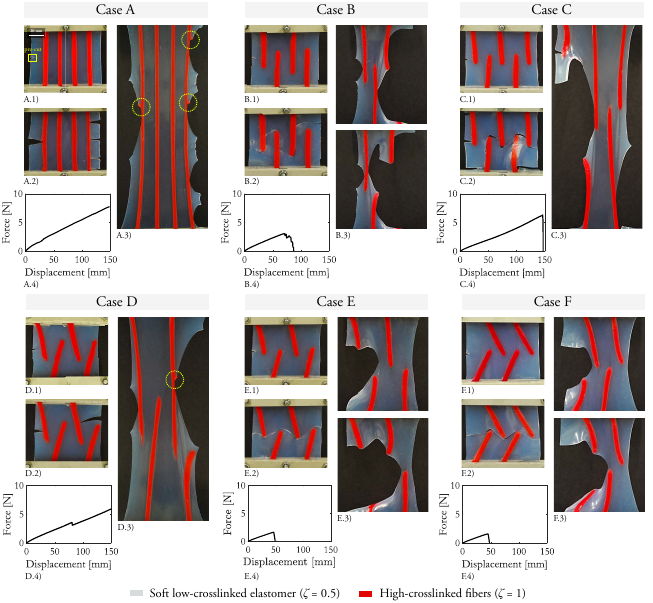}
\caption{\textbf{Composite structures with highly-crosslinked reinforcements to control crack propagation.} A low-crosslinked phase ($\zeta=0.5$) is combined with a highly-crosslinked phase ($\zeta=1$) to produce structures able to control fracture. The highly-crosslinked phase is inserted in the material as reinforcement fibers according to six alternative distributions, denoted as Cases A to F (A-F). Case A and D arrest crack propagation, while Case B, C, E, and F modify the crack path up to eventual final rupture of the low-crosslinked phase. For Cases A and D, sideways crack are marked on the images for the deformed configuration. Images in the undeformed configuration of the samples before crack onset (A-F.1) and after crack propagation (A-F.2). Images in the deformed configuration are included after crack propagation (A-F.3). The force-displacement curves measured experimentally quantify the displacement at failure for the samples undergoing full rupture and the effective stiffness of the samples (A-F.4). For visualization purposes, the highly-crosslinked fibers are marked in red color with a coloring additive. }
\label{fig:composite_frac}
\end{figure}

The smart designs we propose allow either to control the crack path until eventual full rupture, delaying crack propagation, or even to arrest fracture and prevent at all rupturing of the structure. We fabricate structures according to six different topology distributions of the fibers and apply pre-cuts of \qty{5}{\milli \meter} on one side (cases from A to F in Fig.~\ref{fig:composite_frac}.A-F.1). The samples are stretched and fracture onset and propagation are monitored in the unloaded and loaded configurations (material and spatial configurations, respectively, in Figs.~\ref{fig:composite_frac}.A-F.2 and \ref{fig:composite_frac}.A-F.3). On the one hand, the designs in Case B, C, E, and F are able to deviate the crack path, which develops as forward cracking in the low-crosslinked phase. Here, larger fracture paths are linked to larger work of fracture, since more fracture surfaces need to be created. This is quantified by the force displacement curves (Figs.~\ref{fig:composite_frac}.A-F.4), in which the displacement at failure is \qty{48}{\milli \meter} for Case F, \qty{49}{\milli \meter} for E, \qty{87}{\milli \meter} for B, and \qty{147}{\milli \meter} for C. On the other hand, Cases A and D arrest crack propagation in the low-crosslinked phase. Subsequently, sideways cracking at larger displacement values initiates in the highly-crosslinked fibers in the loading direction. The reader is already aware that this type of cracking produces an abrupt increase in the fracture resistance, as described in the previous Fig.~\ref{fig:comput_frac}.

The effective stiffness of the structures varies with the amount and disposition of the highly-crosslinked fibers. As showcased in Fig.~\ref{fig:Exp_mixing}, the quasi-static Young's modulus for low- and high-crosslinked elastomer ($\zeta=0.5$ and \qty{1}) is \qty{10.1}{\kilo \pascal} and \qty{139.3}{\kilo \pascal}, respectively. The difference in more than one order of magnitude and the strong geometrical non-linearity in the deformation produce a highly non-linear constitutive behavior, difficult to predict and understand without a computational continuum model.

%%% FIGURE 4

\subsection{Virtual testing framework to program \& control fracture via composite elastomers}
The previous experimental results demonstrate that mixing low and high crosslinked elastomers provide, at once, low effective stiffness and high fracture toughness. Additive manufacturing techniques can enable the fabrication of soft structures where the highly-crosslinked reinforcements are distributed in a more complicated fashion. In this regard, the three-dimensional distribution of the reinforcements and the creation of intricate, low-scale geometries poses additional challenges. Since the fabrication and experimentation with such designs is not straightforward, we demonstrate that our modeling framework serves as a virtual testbed to assist the design of such structures. 

\begin{figure}[h!]
\centering
\includegraphics[width=1\textwidth]{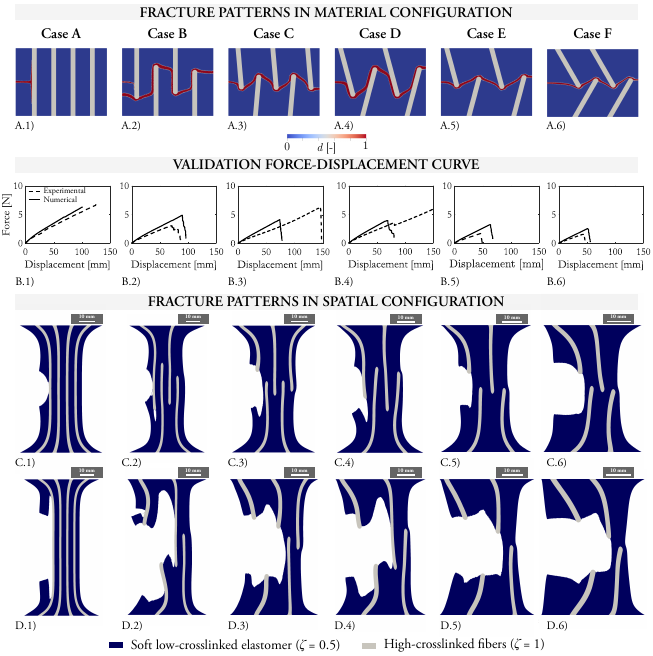}
\caption{\textbf{Validation of the numerical framework for fracture of composite structures.} Composite samples with low-crosslinked ($\zeta=0.5$) and high-crosslinked ($\zeta=1$) phases are modeled to mimic the designs and loading conditions in Fig.~\ref{fig:composite_frac}. (A.1-6) Images in the undeformed configuration of the samples and after crack propagation for Cases A to F. (B.1-6) Comparison of the numerical and experimental force-displacement curves. (C.1-6) Images in the deformed configuration during the tensile deformation of the samples, for Cases A to F, for displacements of the virtual clamps of \qty{76.6}{\milli \meter}, \qty{90}{\milli \meter}, \qty{74.5}{\milli \meter}, \qty{69}{\milli \meter}, \qty{64}{\milli \meter}, and \qty{53}{\milli \meter}, respectively. 
(D.1-6) Images in the deformed configuration during the tensile deformation of the samples, for Cases A to F, for displacements of the virtual clamps of \qty{101}{\milli \meter}, \qty{94}{\milli \meter}, \qty{76}{\milli \meter}, \qty{73.5}{\milli \meter}, \qty{64.5}{\milli \meter}, and \qty{54.6}{\milli \meter}, respectively. A threshold in the damage order parameter of $d=0.5$ is used to removed the damaged areas from the visualization.}
\label{fig:composite_frac_comput}
\end{figure}

As a preliminary step, we demonstrate that our modeling framework is able to replicate the experimental results for the composite structures in Fig.~\ref{fig:composite_frac}. The crack paths obtained from the simulations, depicted in Fig.~\ref{fig:composite_frac_comput}.A.1-6, match the experimental findings. The fracture mechanism relates to fracture of the low-crosslinked phase, eventually propagating along the interface between the low- and high-crosslinked phases. The sideways cracking mechanism as described in Fig.~\ref{fig:comput_frac} occurs only for even larger displacements. Note that only Case D predicts full rupture while the experimental counterpart managed to arrest crack propagation. We believe that fracture propagation at the interface between the phases in the experimental tests is reasonably sensitive to experimental imperfections in the manufacturing process. This fact, together with the experimental deviations when mounting the samples inside the clamps, may justify the disagreement between the experimental and numerical force-displacement curves. For this case, the virtual testbed provides consistent results untethered from such empirical inaccuracies. The physical motivation of all parameters in the model, as detailed in Supplementary Material (Results 1), endorse the reliability of the simulations. The displacement at failure predicted by the model matches the experimental results with relative errors of \qty{9}{\%} for Case B, \qty{47}{\%} for Case C, \qty{28}{\%} for Case E, and \qty{15}{\%} for Case F. We emphasize the convincingly good agreement between the numerical and experimental crack patterns.

%%% FIGURE 5

To conclude the study and illustrate the capabilities of the virtual testbed, we explore the use of simple composite geometries as fracture toughness enhancers, yet preserving the effective softness of the composite structure. The framework is able to guide the design of more intricate topologies with specific constitutive and fracture behavior, which may be integrated in the future to produced tessellated structures. Here, the use of inclusions embedded in the soft matrix Representative Volume Element (RVE) is an interesting case study. We define six three-dimensional RVEs of the mesostructural composition subject to free stress and fixed tensile displacement boundary conditions. One is a pure low-cross crosslinked cube that serves as a baseline for comparison of the other cases. Two cases model high-crosslinked spherical and cubic inclusions in the low-crosslinked matrix. Two others include high-crosslinked cylindrical and prismatic fibers embedded in the middle of the soft cube, with their extreme surfaces in the Dirichlet boundary. 
The sixth RVE models a more complex case with a C-shaped inclusion able to arrest fracture in the soft phase and subsequently developing sideways fracturing inside the high-crosslinked reinforcement. 

\begin{figure}[h!]
\centering
\includegraphics[width=1\textwidth]{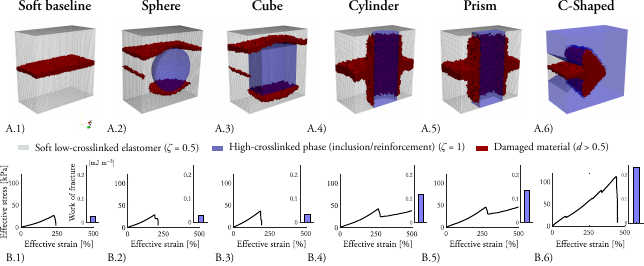}
\caption{\textbf{Results for the fracture behavior of Representative Volume Elements with high-crosslinked inclusions ($\zeta=1$) embedded in a low-crosslinked soft matrix ($\zeta=0.5$).} The \textit{x}-displacement is unconstrained and the \textit{z}-displacement is constrained to zero only at the $x-y$ symmetry plane. Initial damage ($d=1$) is prescribed on one side at the middle of the height. Computational domain in the material (undeformed) configuration where fracture has evolved ($d>0.5$) for a (A.2) spherical inclusion, (A.3) cylindrical fiber, (A.4) cubic inclusion, (A.5) prismatic fiber, and (A.6) C-shaped inclusion. (A.1) A purely low-crosslinked RVE is included as a baseline for comparison to the other cases. The length of the edges of the cubic RVEs is \qty{10}{\milli \meter}, the radius of the centered spherical inclusion \qty{3}{\milli \meter}, \qty{2}{\milli \meter} diameter for the cylindrical fiber, the edge of the cubic inclusion \qty{6}{\milli \meter}, and \qty{4}{\milli \meter} for the prismatic fiber. (B.1-6) Curves with the engineering average stress against the engineering average strain for all cases, including the work of fracture as the area under the curves divided by the total volume of the RVE. For the Cylinder and Prism fiber cases, the work of fracture is calculated up to a displacement of \qty{50}{\milli \meter}.}
\label{fig:composite_predictive}
\end{figure}

Fracture with the RVEs may occur in three different ways: forwards propagation within the low-crosslinked phase, propagation along the interface of the low- and high-crosslinked phases, or sideways propagation within the high-crosslinked phased (see Fig.~\ref{fig:composite_predictive}.A.1-6). The plots with the effective stress and strain of the RVE, together with the information on the work of fracture in Fig.~\ref{fig:composite_predictive}.B.1-6, provide a measure of the effective stiffness before fracture onset and during its propagation. For the purely low-crosslinked, sphere inclusion, and cube inclusion RVEs, the strains at fracture are \qty{241}{\%}, \qty{213}{\%}, and \qty{195}{\%}, respectively. For the cylinder and prism reinforcements, the soft matrix ruptures but the reinforcement provides enhanced fracture resistance. 
The C-shaped design delays the propagation of the fracture with eventual sideways cracking of the high-crosslinked phase, followed by full rupture. To illustrate the mechanism, we refer to the Supplementary Material (Video 1). 
Furthermore, we note that the boundary conditions applied to the RVE determine its response. In this regard, alternative ways to load it may better mimic the integration of the unit cell in a macroscopic, tessellated arrangement. For the sake of illustration, in the Supplementary Material (Results 5) we present the same investigations but for plane strain conditions constraining to zero the z-displacement. From these results, the difference between plane stress and plane strain conditions is convincingly negligible. Overall, the crack-deviatory mechanism of the fracture in the low-crosslinked phase and the sideways propagation mechanism within the highly-crosslinked phase, as well as their simultaneous use within a structure, open numerous possibilities to design fracture in soft structures.

%%% FIGURE 6

\section{Discussion}
Fracture in elastomers can manifest through unconventional patterns, particularly in highly crosslinked elastomers where sideways fracturing significantly enhances fracture toughness. In this work, we propose a roadmap to alter crack paths and drastically improve fracture toughness. Through comprehensive experimental characterization and an advanced phase-field numerical model, we explore the formation of sideways cracking patterns. We demonstrate that the mechanism driving sideways cracking is twofold: geometrically nonlinear fracture and anisotropic increase in the critical energy release rate in the direction perpendicular to the highly stretched polymer chains. Our findings show that increasing the crosslinking degree to an optimal level drastically enhances the critical fracture energy release rate via sideways cracking. To capture this strain-induced fracture anisotropy in a theoretical description, we propose a novel phase-field constitutive model for sideways cracking that directly relates the critical fracture energy release rate to the deformation at the crack tip front. This model provides a deeper understanding of the anisotropic toughening mechanisms and offers a new approach for designing elastomers with superior fracture resistance.

To integrate the advantageous low stiffness of low-crosslinked elastomers with the fracture-arresting capabilities of high-crosslinked elastomers, we extended our computational framework to function as a virtual testbed for designing composite structures. By incorporating high-crosslinked fiber reinforcements within the matrix, we demonstrated that crack propagation can be delayed and even arrested. The computational framework successfully replicated crack patterns in a simplified two-dimensional context and remains capable of predicting fracture behavior in more complex composite structures. Using this framework as a standalone tool, we conducted simulations on three-dimensional RVEs to understand the fracture performance of mesostructures that incorporate high-crosslinked inclusions within soft low-crosslinked matrices. The combined use of the crack-deviating mechanism along the interface between low- and high-crosslinked phases and the sideways propagation mechanism within the high-crosslinked phase opens numerous possibilities for controlling fracture in soft structures. This innovative approach enables the design of materials that combine the best properties of both phases, offering new pathways for creating soft, yet ultra-tough, materials with enhanced fracture resistance.

Our framework will enable predictive capabilities and even the creation of databases for fracture. Information generated from our model, such as the work of fracture of a minimal composite cell and its uniform stiffness, as illustrated in Fig.~\ref{fig:composite_predictive}, can inform machine learning databases that optimize the fracture performance under specific design conditions. Taking the inspiration of model-driven identification frameworks, we envision a machine-learning-based approach informed by a library of minimal cells exhibiting diverse constitutive and fracture behaviors. By strategically arranging these cells, potentially in a metastructure configuration, it becomes feasible to achieve desired fracture behaviors. This integrated approach not only enhances our understanding of fracture mechanics but also opens new avenues for designing materials with tailored fracture resistance and performance characteristics.

Furthermore, data-driven methods hold promise for constructing more accurate constitutive models for soft materials with highly complex material responses, as described in previous works by the authors \cite{Wiesheier2023,Wiesheier2024}, and even for aiding in calibrating fracture behavior. Among other strategies, topology optimization and fracture optimal control may also enable smart fracture programming \cite{Martinez-Frutos2021c,Khimin2022}. Such advancements may inspire further fracture studies in hydrogels and other multiphase systems aimed at controlling fracture, as evidenced by recent research \cite{Xiao2022}. This progress opens doors to engineering applications such as cutting-edge fracture mechanics in computer-guided surgery \cite{Courtecuisse2014}, where precise control over material fracture properties is crucial for enhancing surgical outcomes and patient safety.

An intriguing point is also the integration of advanced theories for fracture, such as configurational mechanics. Here we highlight the recent application of the authors of the Configurational Force Method to fracture in soft materials undergoing large geometrical non-linearities \cite{Moreno-Mateos2024b}. The application of configurational forces to analyze soft fracture establishes a promising framework for studying nonlinear fracture. When coupled with sideways cracking, this approach opens up new avenues for developing theoretical models that can accurately capture stretch-induced anisotropic fracture in soft materials.

\section{Methods}
\subsection{Materials and synthesis of single-material samples}\label{sec:materials_synthesis}
%%%%%%%%%%%%%%%%%%%%%%%%%%
We manufacture samples from the elastomer Elastosil P7670 (Wacker Chemie AG, Munich, Germany). The material is provided as two raw phases (phase A and B) so that the platinum catalyst is separated from the curing agent. After the components are mixed, the mixture cures to form a crosslinked silicone elastomer. The mixing ratio of the phases is defined through the parameter crosslinking ratio $\zeta=V_\text{B}/V_\text{A}$, with $V_\text{B}$ and $V_\text{A}$ the volume of the phase A and B, respectively. To adjust the crosslinking degree, we modify the mixing ratio according to $\zeta=\{0.5, 0.625, 0.75, 0.875, 1,1.14, 1.33, 1.6, 2 \}$. In this way, and not needing additional crosslinker substances, materials with different material response are synthesized. The samples are cured during \qty{2}{\hour} in an open mold at a temperature of \qty{120}{\celsius}. The thickness ($t$) of all samples is 2 mm and they are die-cut to produce rectangles with a width ($w$) of $12~\rm{mm}$. Furthermore, two types of samples are defined: pristine (or uncut) samples and pre-cut samples. For the latter ones, we apply pre-cuts with a razor blade and lengths ($c$) of \qty{1.2}{\milli \meter}, \qty{3.6}{\milli \meter}, and \qty{6}{\milli \meter}. These define crack-width ratios ($c/w$) of \qty{0.1}, \qty{0.3}, and \qty{0.5}, respectively. More details on the manufacturing procedure are provided in the Supplementary Material (Methods 1).
%%%%%%%%%%%%%

\subsection{Experimental testing of single-material samples}\label{sec:experimental}
An universal tensile machine (Inspekt S 5 kN, Hegewald \& Peschke, Nossen, Germany) was used to perform tensile tests on pristine and pre-cut samples with an initial length ($l_0$) of \qty{30}{\milli \meter} and with loading rates of $0.03$~\unit{\mm \per \second} and $3$~\unit{\mm \per \second}, which render average strain rates, respectively, of $0.001$~\unit{\per \second} (quasi-static) and $0.1$~\unit{\per \second}. The clamps on the machine were actuated with air pressure. The force-displacement data was stored during the deformation of the samples until fully rupture or, alternatively, until slipping between the sample and the grips for the higher mixing ratios, i.e., higher crosslinking degrees and stiffer samples. Simultaneously, a monochromatic CCD sensor (DCS 2.0, LIMESS Messtechnik \& Software GmbH, Germany) with a resolution of 1024x768 and a lens with focal range \qty{50}{\milli \meter} and aperture 2.8-16 (2.8/50-0902 Xenoplan, Schneider Kreuznach, Bad Kreuznach, Germany) were used to capture the cracking pattern.

\subsection{Materials and synthesis of composite samples}\label{sec:materials_synthesis_composite}
Composite samples to control fracture were manufactured combining the same structure elastomers with different mixing ratios. The blend with high mixing ratio for the highly-crosslinked reinforcement was poured in a container and a red colorant was added. Then, a centrifuge was used to homogenize the blend during \qty{2}{\min} at 3000~RPM. Then, the blend was injected on the mold with the help of an auxiliary 3D-printed mold according to the desired geometry of the highly-crosslinked reinforcement. A Ultimaker 2+ printer (Ultimaker, Utrecht, Netherlands) was used to prepare the molds. A syringe was used to fill the auxiliary mold. The polymer was pre-cured afterwards for \qty{15}{\min} at room temperature (\qty{25}{\celsius}), after which the auxiliary mold was removed. The soft elastomer blend ($\zeta=0.5$), immediately before prepared with \qty{1}{\min} in the centrifuge, was added to fill the gap between the pre-cured stiff polymer. Both blends in touch were finally cured during \qty{2}{\hour} in an oven at \qty{120}{\celsius}. The resulting samples have a thickness of \qty{2}{\milli \meter} and width of \qty{95}{\milli \meter}. More details on the manufacturing procedure are provided in the Supplementary Material (Methods 1).

\subsection{Experimental testing of composite samples}\label{sec:experimental_composite}
The composite samples are tested with an initial length ($l_0$) of \qty{70}{\milli \meter} and with a loading rate of \qty{0.07}{\milli \meter \per \second}. 
The universal tensile machine (Inspekt S 5 kN, Hegewald \& Peschke, Nossen, Germany) was used to test the samples under tensile loading up to full rupture or \qty{150}{\milli \meter} in the case of samples able to arrest crack propagation. Simultaneously, a Sony alpha 6700 camera was utilized to acquire images of the sample during the loading process, with shutter speed 1/25, aperture 4.5, and ISO 400.

%%%%%%%%%%%%%%%%%%%%%%%%%%
\subsection{Constitutive and computational phase-field framework for sideways cracking}\label{sec:constitutive_framework}
%%%%%%%%%%%%%%%%%%%%%%%%%%
The deformation of the medium is formulated in a finite strain framework. The primary fields to be solved are the displacement field vector $\mathbf{u}$ and the scalar order parameter $d$. The deformation gradient is defined as $\mathbf{F}=\nabla_0\mathbf{u} + \mathbf{I}$, with $\mathbf{I}$ the second-order identity tensor and $\nabla_0$ the gradient operator in the material configuration. Following the multiplicative isochoric-volumetric decomposition $\mathbf{F}=\mathbf{F}_\text{vol}\cdot\overline{\mathbf{F}}$, the volumetric part is defined as $\mathbf{F}_\text{vol}=J^{1/3}\mathbf{I}$ and the isochoric part as $\overline{\mathbf{F}}=J^{-1/3}\mathbf{F}$, where $J=\det \mathbf{F}$ denotes the determinant of $\mathbf{F}$.

The first variation of the incremental rate-dependent functional in Equation~S.1 (Supplementary Material (Methods 2)) renders the  localized quasi-static balance field equation as
\begin{equation}\label{eq:fieldeq_mech}
\nabla_0 \cdot\mathbf{P} = \mathbf{0},
\end{equation}
\noindent where $\mathbf{P}$ refers to the Piola stress tensor.

Likewise, the phase-field equation for fracture results
\begin{equation}\label{eq:fieldeq_damage}
   G_\text{c}\left(\mathbf{F},d\right) \frac{3}{8} \left[\frac{1}{l} - 2 l \nabla_0^2 d \right]+g'\left(d\right) \Psi +\eta\dot{d}
  = 0,
\end{equation}
\noindent with $g\left(d\right):=\left[1-d\right]^2$ the degradation function and $g'\left(d\right)=2\left[d-1\right]$ its derivative with respect to $d$. Moreover, $l$ is the characteristic length in the diffusive crack topology, approaching the discrete crack topology as $l\rightarrow0$. The regularization of the crack in Equation~\ref{eq:fieldeq_damage} corresponds to Bourdin, Francfort, and Marigo \cite{Bourdin2000,Bourdin2007,Francfort2008}. The crack surface density, a geometrical quantity, is defined following the Ambrosio and Tortorelli \texttt{AT-1} model \cite{Ambrosio1990}, as detailed in Equation~S.3 (Supplementary Material (Methods 2)). Further, the scalar $\eta$ denotes a viscosity parameter to enhance the numerical robustness of the damage field evolution and $\Psi$ the total energy density driving fracture propagation. 

The critical energy release rate ($G_\text{c}$) defined in Equation~\ref{eq:fieldeq_damage} sets the energy threshold for crack initiation. To model anisotropic cracking in an isotropic continuum (i.e., with isotropic $\Psi$), we propose a direction-dependent material function $G_\text{c}=G_\text{c}\left(\mathbf{F},d\right)$ comprising isotropic ($G_\text{c,iso}$) and anisotropic ($G_\text{c,ani}$) contributions. The former ($G_\text{c,iso}$) remains constant and represents the resistance to fracture in conventional forward cracking. We estimate a value using the samples with low crosslinking degree ($\zeta=0.5$) as the energy release between two specimens with precuts of different lengths (see more detail in the Supplementary Material (Results 2)). The latter ($G_\text{c,ani}$) varies with the propagation direction, as it differs when propagating by cutting polymer chains compared to propagation parallel to the chains, and depends on the stretch in the direction of the chains. Consequently, the energetic requirements increase significantly for a crack propagating perpendicular to the chains aligned in the loading direction and experiencing severe stretching. The total critical energy release rate is eventually given by
%\begin{equation}\label{eq:gamma_crack_regularisation}
%G_\text{c} \left(\mathbf{F},d,\hat{\mathbf{n}}_{\text{ani},\perp}\right) = 
%G_\text{c,iso} + 
%\underbrace{\beta_\text{ani} \left(\mathbf{F},d\right)  \left[\hat{\mathbf{n}}_{\text{ani},\perp}  \cdot \hat{\boldsymbol \nabla d}\right]}_{G_\text{c,ani}}.
%\end{equation}
\begin{equation}\label{eq:gamma_crack_regularisation}
G_\text{c} \left(\mathbf{F},d\right) = 
G_\text{c,iso} + 
\underbrace{\beta_\text{ani} \left(\mathbf{F},d\right)  ||\widehat{ \nabla} d _{\perp \hat{\mathbf{m}}_\text{ani}}||}_{G_\text{c,ani}\left(\mathbf{F},d\right)}.
\end{equation}

The orientation of the polymer chains (crystal domains) is determined in the spatial configuration by the boundary value problem. We denote $\mathbf{m}_\text{ani}=c \, \hat{\mathbf{m}}_\text{ani}$ for $\hat{\mathbf{m}}_\text{ani}$ a normal vector and $c$ an unknown scalar. In experiments with uniaxial load, such as ours, $\mathbf{m}_\text{ani}$ corresponds to the loading direction in the spatial configuration, e.g., defined through a vertical  $\hat{\mathbf{m}}_\text{ani}$ unit vector (see Supplementary Material (Figure S5)). Note that only the direction of the $\mathbf{m}_\text{ani}$ vector, and not its norm, is required. This direction defines a perpendicular plane $\rm{P}$, with $\hat{\mathbf{m}}_\text{ani}$ its normal vector. Thus, the vector rejection of the crack direction vector from the plane's normal vector, i.e., the projection on the plane, can be calculated as $\widehat{ \nabla }d _{\perp \hat{\mathbf{m}}_\text{ani}}=\left[\hat{\mathbf{m}}_{\text{ani}}  \times \widehat{\nabla} d\right] \times \hat{\mathbf{m}}_{\text{ani}} =  \widehat{ \nabla} d - \left[\widehat{ \nabla} d \cdot \hat{\mathbf{m}}_\text{ani}\right]\hat{\mathbf{m}}_\text{ani}$. 

For general problems, $\hat{\mathbf{m}}_\text{ani}$ can be related to the principal loading direction. In turn, the material gradient of the damage field can be expressed in terms of the material gradient as $\nabla d=\nabla_0 d \cdot \mathbf{F}^{-1}$, thus allowing us to calculate $\widehat{\nabla} d=\nabla d/\left[k+|\nabla d|\right]$ as the normalized unit gradient vector. To prevent singularity in its normalization when $\nabla d$ approaches zero, we add a small parameter $k=\qty{1e-6}{}/l$ to the denominator.

The amplification function $\beta_\text{ani} \left(\mathbf{F},d\right)$ in Equation~\ref{eq:gamma_crack_regularisation} allows for mimicking the increase of fracture anisotropy with the deformation of the chains. For an undeformed elastomer, $\beta_\text{ani}\left(\mathbf{F}=\mathbf{I},d\right)=0$ and $G_\text{c}$ recovers the isotropic critical energy release rate. Let us define a potential relation with the strain in the loading direction according to
\begin{equation}\label{eq:beta_aniso}
\beta_\text{ani}\left(\mathbf{F},d\right) = \tilde{\beta}_\text{ani} \, G_\text{c,iso} \, g\left(d\right) \left\langle \mathbf{E}:\left[\hat{\mathbf{M}}_\text{ani}\otimes\hat{\mathbf{M}}_\text{ani}\right]\right\rangle_+^\alpha.
\end{equation}

In Equation~\ref{eq:beta_aniso}, the \textit{Green-Lagrange} strain tensor $\mathbf{E}=\frac{1}{2}\left[\mathbf{\mathbf{F}^\text{T}\cdot\mathbf{F}}-	\mathbf{I}\right]$, a covariant tensor, is double-contracted with the normal chains direction in the material configuration, $\hat{\mathbf{M}}_\text{ani}$, a tangent vector obtained through pull-back (contravariant) operation of the spatial counterpart, i.e., $\hat{\mathbf{M}}_\text{ani} = \mathbf{F}^\text{-1}\cdot \left[c \, \hat{\mathbf{m}}_\text{ani}\right]$, with $c$ determined so that $\hat{\mathbf{M}}_\text{ani}$ is unit vector. This allows to make the amplification function directly dependent on the strain of the chains in the loading direction. Note that the positive ramp function $\langle \bullet \rangle_+$ prevents amplification of $G_\text{c}$ under compression, but only under tensile loading of the chains.  Furthermore, the factor $\tilde{\beta}_\text{ani}$ and the exponent $\alpha$ are heuristic parameters that establish how the anisotropic critical energy release rate scales with the stretch of the polymer chains. In the present work, we select $\alpha=1$. Eventually, the use of the degradation function in the former equation is necessary to decrease the anisotropic fracture resistance once damage evolves. Otherwise, the very high deformation of damaged elements leads to nonphysical values of $G_\text{c}$, entailing potential damage reversibility. We note that for more complex loading modes Equation~\ref{eq:beta_aniso} may be modified to not only depend on the principal loading direction $\hat{\mathbf{m}}_\text{ani}$, but on a more representative measure of the deformation state. This may constitute a point for future work.

The total energy density, which decomposes into isochoric ($\Psi_\text{iso}$) and volumetric ($\Psi_\text{vol}$) contributions, is degraded with the damage parameter according to
\begin{equation}\label{eq:total_energy_density}
\Psi = g\left(d\right) \left[ \Psi_\text{iso} + \Psi_\text{vol} \right].
\end{equation}

The isochoric contribution to the total energy density in Equation~\ref{eq:total_energy_density} is defined according to the Yeoh model as
\begin{equation}\label{eq:Psi_isochoric}
\Psi_\text{iso}\left(\overline{\mathbf{F}}\right)=
C_{1}\left[\overline{I}_1-3\right]+C_{2}\left[\overline{I}_1-3\right]^2+C_{3}\left[\overline{I}_1-3\right]^3,
\end{equation}
\noindent with the isochoric invariant $\overline{I}_1=\text{tr} \left(\overline{\mathbf{F}}^\text{T}\cdot\overline{\mathbf{F}}\right)$ and $2C_{1}$ the shear modulus. The calibration of the coefficients $C_{1}$, $C_{2}$, and $C_{3}$ parameters is detailed in the Supplementary Material (Results 1).

For the volumetric contribution, we use a relation directly dependent on the bulk modulus that is adequate to recover the nearly incompressible behavior of elastomers, i.e.,
\begin{equation}
\Psi_{\text{vol}} \left( J \right) =
\frac{\kappa}{2} \left[J-1\right]^2,
\quad 
\textrm{with}
\quad
\kappa=\frac{4C_1\left[1+\nu \right]}{3\left[1-2\nu\right]},
\end{equation}
\noindent for bulk modulus $\kappa$ with Poisson ratio $\nu$ set to 0.49.

Subsequently, the Piola stress tensor can be derived from the energy density as the addition of isochoric and volumetric contributions according to $\mathbf{P}=g\left(d\right)\mathbf{P}_\text{iso}+g\left(d\right)\mathbf{P}_\text{vol}$, via
\begin{align}\label{eq:Piola_stress}
\mathbf{P} = \frac{\partial \Psi \left(\mathbf{F}\right)}{\partial \mathbf{F}}.
\end{align}

The complete constitutive framework is presented in the Supplementary Material (Methods 2).

%\subsection{Numerical implementation}
For the numerical implementation of the field equations, their weak forms need to be derived testing and integrating the field equations, as detailed in Equations~S.12 and S.13 in the Supplementary Material (Methods 3). The weak form of the problem, which is numerically solved in the open-source finite element environment FEniCS. The finite element computations were performed on a two-dimensional mesh with quadratic triangular elements. The mesh replicates only the upper half of the samples due to horizontal symmetry. In this regard, we note that sideways cracks do not occur strictly in a symmetric fashion. Nonetheless, we observe that a second sideways crack symmetric to the initial one initiates once the first one stops propagating, and in certain experiments the onset of both cracks occurs simultaneously. With this simplified modeling assumption, we preserve the physical behavior while reducing the computational expenses for finer FE meshes. To mimic a tensile test, the displacement of the upper edge of the rectangular sample is constrained, whereby the vertical displacement is prescribed.

While 3D simulations are conducted to calibrate the constitutive parameters, see the Supplementary Material (Results 1), 2D simulations are performed for the fracture simulations in Section~\ref{sec:comput_model_sideways}. This allows to reach smaller mesh sizes than with a 3D mesh, while keeping the computational expenses within computation limits and thus the computation times efficiently bounded. The effect of the thickness of the samples on the sideways cracking mechanism is not object of the present study. The reader interested in such effect can consult \cite{Lee2019c}. According to this work, the limited thickness of the samples promotes forward cracking. Although this effect is not captured with our 2D simulations, the error due to this modeling simplification is deemed insignificant and does not hinder the qualitative and quantitative fracture behavior. 

More details on the numerical implementation of the computational model are given in Supplementary Material (Methods 3).

\section*{Data and Software Availability}
%The data generated during the study is available via zenodo in XX. 
The data generated during the study will be made available via zenodo upon acceptance of the manuscript.

\section*{Code Availability statement}
\noindent The code generated during the current study is available from the corresponding author upon reasonable request.

\section*{Acknowledgments}
The authors acknowledge support from the European Research Council (ERC) under the Horizon Europe program (Grant-No. 101052785, project: SoftFrac).

\section*{Competing Interests}
\noindent The Authors declare no Competing Financial or Non-Financial Interests.

\section*{Supporting Information Appendix (SI)}
The online version contains Supplementary Material

\newpage

{
\captionsetup[figure]{labelsep=period}
\def\vc #1{\mbox{\boldmath $#1$}}
\renewcommand{\figurename}{Figure}
\renewcommand{\thefigure}{S\arabic{figure}}
\captionsetup{labelsep = period}

\renewcommand{\tablename}{Table}
\renewcommand{\thetable}{S\arabic{figure}}
\renewcommand{\theequation}{S.\arabic{equation}}
\setcounter{figure}{0}
\setcounter{equation}{0}

\vspace{5cm}

\begin{center}
{\huge SUPPLEMENTARY MATERIAL}\\
\vspace{0.5cm}
{\LARGE Crosslinking degree variations enable programming and controlling soft fracture via sideways cracking}
\end{center}

\newpage

\subsection*{\textup{\textbf{Methods 1: Synthesis of elastomer samples: single-material rectangular samples and composite samples}}}
Single-material samples are prepared with the elastomer Elastosil P7670 (Wacker Chemie AG, Munich, Germany). The material is provided as two raw phases (phase A and B). After the components are mixed, the mixture cures to form a crosslinked silicone elastomer. The synthesis methodology is depicted in Figure~\ref{fig:Manuf}.A. The mixing ratio of the phases is defined through the parameter crosslinking ratio $\zeta=V_\text{B}/V_\text{A}$, with $V_\text{b}$ and $V_\text{a}$ the volume of the phase A and B, respectively. To adjust the crosslinking degree, we modify the mixing ratio according to $\zeta=\{0.5, 0.625, 0.75, 0.875, 1,1.14, 1.33, 1.6, 2 \}$. In this way, and not needing additional crosslinker substances, materials with different material response are synthesized. The samples are degassed in a vacuum chamber for \qty{10}{\min} and then cured during \qty{2}{\hour} in an open mold at a temperature of \qty{120}{\celsius}. The thickness ($t$) of all samples is 2 mm and they are die-cut to produce rectangles with a width ($w$) of $12~\rm{mm}$. Furthermore, two types of samples are defined: pristine (or uncut) samples and pre-cut samples. For the latter ones, we apply pre-cuts with a razor blade and lengths ($c$) of \qty{1.2}{\milli \meter}, \qty{3.6}{\milli \meter}, and \qty{6}{\milli \meter}. These define crack-width ratios ($c/w$) of \qty{0.1}, \qty{0.3}, and \qty{0.5}, respectively.

Fiber-reinforced composite samples were manufactured combining the same structure elastomers with different mixing ratios. The synthesis methodology is depicted in Figure~\ref{fig:Manuf}.B. The blend with high mixing ratio for the highly-crosslinked reinforcement was poured in a container. One drop of red colorant was added to the last one. Then, a centrifuge was used to homogenize the blend during \qty{2}{\min} at 3000~RPM. Then, the blend was injected on the mold with the help of an auxiliary 3D-printed mold according to the desired geometry of the highly-crosslinked reinforcement. A Ultimaker 2+ printer (Ultimaker, Utrecht, Netherlands) was used to prepare the molds. A syringe was used to fill the auxiliary mold. The polymer was pre-cured afterwards for \qty{15}{\min} at room temperature (\qty{25}{\celsius}), after which the auxiliary mold was removed. The soft elastomer blend ($\zeta=0.5$), immediately prepared before with \qty{1}{\min} in the centrifuge, was added to fill the gap between the pre-cured stiff polymer. Both blends in touch were finally cured during \qty{2}{\hour} in an oven at \qty{120}{\celsius}. The resulting samples have a thickness of \qty{2}{\milli \meter} and width of \qty{95}{\milli \meter}.

\begin{figure}
\centering
\includegraphics[width=1\textwidth]{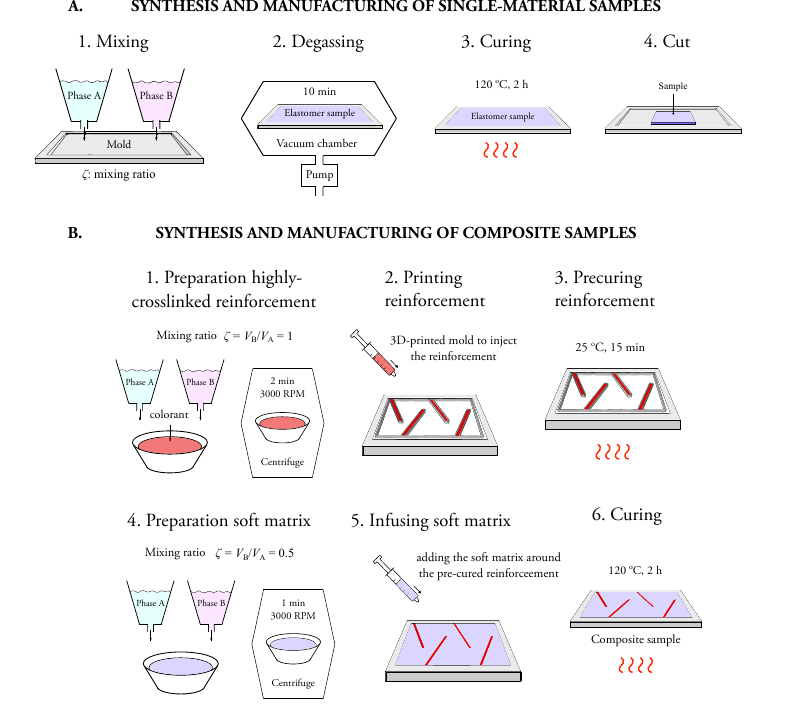}
\caption{\textbf{Overview of the methodology to manufacture elastomer samples with different crosslinking degrees and composite samples with high-crosslinked reinforcements.} (A) Synthesis of single-material samples with an open mold. First, the two raw phases (phase A and B) are mixed attending to the volume mixing ratio $\zeta=V_\text{b}/V_\text{a}=\{0.5, 0.625, 0.75, 0.875, 1,1.14, 1.33, 1.6, 2 \}$, with $V_\text{B}$ and $V_\text{A}$ the volume of the phase A and B, respectively. Second, the samples are degassed in a vacuum chamber for \qty{10}{\min}. Third, the samples are cured in an open mold at \qty{120}{\celsius} during \qty{2}{\hour}. Fourth, the sames are die-cut to produce rectangles with a width ($w$) of $12~\rm{mm}$, with a thickness ($t$) of all samples is 2 mm. Two types of samples are defined: pristine (or uncut) samples and pre-cut samples. 
(B) Synthesis of composite samples with an open mold and combining within the same structure elastomers with different mixing ratios. First, the blend with high mixing ratio for the highly-crosslinked reinforcement is poured in a container and a red colorant is added. A centrifuge is used to homogenize the blend during \qty{2}{\min} at 3000~RPM. Second, the blend is injected on the mold with the help of an auxiliary 3D-printed mold. Third, the polymer is pre-cured for \qty{15}{\min} at room temperature (\qty{25}{\celsius}), after which the auxiliary mold was removed. Fourth, the soft elastomer blend ($\zeta=0.5$) is prepared during \qty{1}{\min} in the centrifuge. Fifth, the soft phase is added to fill the gap between the pre-cured stiff polymer. Sixth, the blends in touch are cured in an oven at \qty{120}{\celsius} during \qty{2}{\hour}. The resulting samples have a thickness of \qty{2}{\milli \meter} and width of \qty{95}{\milli \meter}. 
A mold made of polytetrafluoroethylene (PTFE) is used.
}
\label{fig:Manuf}
\end{figure}

\newpage
\subsection*{\textup{\textbf{Methods 2: Complete theoretical framework and constitutive formulation for sideways fracture}}}\label{sec:constit_framework_S}
The deformation of the medium is formulated in a finite strain framework. The displacement field $\mathbf{u}\left(\mathbf{X}\right)$ maps the positions in the material configuration $\mathbf{X} \in \Omega_0$, with boundary $\partial \Omega_0$, to the positions in the spatial configuration $\mathbf{x} \in \Omega$,  with boundary $\partial \Omega$, according to $\mathbf{x}=\boldsymbol \varphi\left(\mathbf{X}\right)=\mathbf{u}\left(\mathbf{X}\right)+\mathbf{X}$. The deformation gradient is defined as $\mathbf{F}=\nabla_0\mathbf{u} + \mathbf{I}$, with $\mathbf{I}$ the second-order identity tensor and $\nabla_0$ the gradient operator in the material configuration. Following the multiplicative isochoric-volumetric decomposition $\mathbf{F}=\mathbf{F}_\text{vol}\cdot\overline{\mathbf{F}}$, the volumetric part is defined as $\mathbf{F}_\text{vol}=J^{1/3}\mathbf{I}$ and the isochoric part as $\overline{\mathbf{F}}=J^{-1/3}\mathbf{F}$, where $J=\det \mathbf{F}$ denotes the determinant of $\mathbf{F}$. 

Damage and degradation of the material properties are modeled through a scalar order parameter $d$. A central notion in the phase-field approach is to model the crack as a diffusive discontinuity where the damage field evolves continuously from zero to one, i.e., from virgin to fully damaged material.

Let $\mathrm{F}$ be an incremental rate-dependent functional in the spirit of the Francfort and Marigo variational setting \cite{Francfort1998}, with internal power ($\dot{\Pi}_\text{int}$), fracture dissipation due to the creation of crack surface ($D_\text{crack}$), viscous fracture dissipation regularization ($D_\text{crack,visc}$), and external power ($\dot{\Pi}_\text{ext}$) due to material traction ($\mathbf{t}_0$) and body forces ($\mathbf{b}_0$),
\begin{align}\label{eq:total_rate_potential_S}
\mathrm{F}=&
\underbrace{\int_{\Omega_0} \left[ \partial_\mathbf{F}\, \Psi : \dot{\mathbf{F}} + \partial_d\Psi\dot d \right] \text{d}V}_{\dot{\Pi}_\text{int}} 
 +
 \underbrace{\int_{\Omega_0} G_\text{c} \left(\mathbf{F},d\right)\frac{3}{8}\left[\frac{\dot{d}}{l} + 2 l \nabla_0 d \cdot   \nabla_0 \dot{d}\right] \text{d}V}_{D_\text{crack}} + 
 \underbrace{\int_{\Omega_0} \frac{\eta}{2}\dot{d}^2 \, \text{d}V}_{D_\text{crack,visc}}
 \\ & 
\underbrace{-\int_{\Omega_0} \, \mathbf{b}_0 \cdot \dot{\mathbf{u}} \, \text{d}V - \int_{\partial {\Omega_0}} \mathbf{t}_0 \cdot \dot{\mathbf{u}} \, \text{d}A }_{\dot{\Pi}_\text{ext}}. \nonumber
\end{align}

The total energy density, which decomposes into isochoric ($\Psi_\text{iso}$) and volumetric ($\Psi_\text{vol}$) contributions, is degraded with the damage parameter according to
\begin{equation}\label{eq:total_energy_density_S}
\Psi = g\left(d\right) \left[ \Psi_\text{iso} + \Psi_\text{vol} \right],
\end{equation}
\noindent with $g\left(d\right):=\left[1-d\right]^2$ the degradation function. 

The regularization of Equation~\ref{eq:total_rate_potential_S} corresponds to Bourdin, Francfort, and Marigo \cite{Bourdin2000,Bourdin2007,Francfort2008}. Following the Ambrosio and Tortorelli \texttt{AT-1} model \cite{Ambrosio1990}, the crack surface density per unit volume ($\gamma$) is a geometrical quantity that can be defined as
\begin{equation}\label{eq:regularized_gamma_S}
\gamma\left(d\right) = \frac{3}{8}\left[\frac{d}{l} +  l  |\nabla_0 d|^2\right],
\end{equation}
\noindent with $l$ the characteristic length in the diffusive crack topology, approaching the discrete crack topology as $l\rightarrow0$. In this work, we select $l=0.05~\mathrm{mm}$ to enable $h<l/2$, with $h$ the element size in the FE mesh. For composite samples, the characteristic length is set to $l=0.1~\mathrm{mm}$. The non-negativity of the damage variable is enforced through an additional penalty term in Equation~\ref{eq:total_rate_potential_S} as, e.g., in \cite{Moreno-Mateos2023} and \cite{Moreno-Mateos2024a}. 

The critical energy release rate ($G_\text{c}$) defined in Equation~\ref{eq:total_rate_potential_S} sets the energy threshold for crack initiation. To model anisotropic cracking in an isotropic continuum (i.e., with isotropic $\Psi$), we propose a direction-dependent material function $G_\text{c}=G_\text{c}\left(\mathbf{F},d\right)$ comprising isotropic ($G_\text{c,iso}$) and anisotropic ($G_\text{c,ani}$) contributions. The former ($G_\text{c,iso}$) remains constant and represents the resistance to fracture in conventional forward cracking. We estimate a value using the samples with low crosslinking degree ($\zeta=0.5$) as the energy release between two specimens with precuts of different lengths (see more detail in Methods 4). The latter ($G_\text{c,ani}$) varies with the propagation direction, as it differs when propagating by cutting polymer chains compared to propagation parallel to the chains, and depends on the stretch in the direction of the chains. Consequently, the energetic requirements increase significantly for a crack propagating perpendicular to the chains aligned in the loading direction and experiencing severe stretching. The total critical energy release rate is eventually given by
\begin{equation}\label{eq:gamma_crack_regularisation_S}
G_\text{c} \left(\mathbf{F},d\right) = 
G_\text{c,iso} + 
\underbrace{\beta_\text{ani} \left(\mathbf{F},d\right)  ||\widehat{ \nabla} d _{\perp \hat{\mathbf{m}}_\text{ani}}||}_{G_\text{c,ani}\left(\mathbf{F},d\right)}.
\end{equation}

The orientation of the polymer chains (crystal domains) is determined in the spatial configuration by the boundary value problem. We denote $\mathbf{m}_\text{ani}=c \, \hat{\mathbf{m}}_\text{ani}$ for $\hat{\mathbf{m}}_\text{ani}$ a normal vector and $c$ an unknown scalar. In experiments with uniaxial load, such as ours, $\mathbf{m}_\text{ani}$ corresponds to the loading direction in the spatial configuration, e.g., defined through a vertical  $\hat{\mathbf{m}}_\text{ani}$ unit vector (see Figure~\ref{fig:exp_cracks_def}). Note that only the direction of the $\mathbf{m}_\text{ani}$ vector, and not its norm, is required. This direction defines a perpendicular plane $\rm{P}$, with $\hat{\mathbf{m}}_\text{ani}$ its normal vector. Thus, the vector rejection of the crack direction vector from the plane's normal vector, i.e., the projection on the plane, can be calculated as $\widehat{ \nabla} d _{\perp \hat{\mathbf{m}}_\text{ani}}=\left[\hat{\mathbf{m}}_{\text{ani}}  \times \widehat{\nabla} d\right] \times \hat{\mathbf{m}}_{\text{ani}} =  \widehat{ \nabla} d - \left[\widehat{ \nabla} d \cdot \hat{\mathbf{m}}_\text{ani}\right]\hat{\mathbf{m}}_\text{ani}$. 

For general problems, $\hat{\mathbf{m}}_\text{ani}$ can be related to the principal loading direction. In turn, the material gradient of the damage field can be expressed in terms of the material gradient as $\nabla d=\nabla_0 d \cdot \mathbf{F}^{-1}$, thus allowing us to calculate $\widehat{\nabla} d=\nabla d/\left[k+|\nabla d|\right]$ as the normalized unit gradient vector. Note that the gradient of the damage parameter has been utilized in the literature to describe anisotropy, as seen in references such as \cite{Kobayashi1993}. To prevent singularity in its normalization when $\nabla d$ approaches zero, we add a small parameter $k=\qty{1e-6}{}/l$ to the denominator.

The amplification function $\beta_\text{ani} \left(\mathbf{F},d\right)$ in Equation~\ref{eq:gamma_crack_regularisation} allows for mimicking the increase of fracture anisotropy with the deformation of the chains. For an undeformed elastomer, $\beta_\text{ani}\left(\mathbf{F}=\mathbf{I},d\right)=0$ and $G_\text{c}$ recovers the isotropic critical energy release rate. Let us define a potential relation with the strain in the loading direction according to
\begin{equation}\label{eq:beta_aniso_S}
\beta_\text{ani}\left(\mathbf{F},d\right) = \tilde{\beta}_\text{ani} \, G_\text{c,iso} \, g\left(d\right) \left\langle \mathbf{E}:\left[\hat{\mathbf{M}}_\text{ani}\otimes\hat{\mathbf{M}}_\text{ani}\right]\right\rangle_+^\alpha.
\end{equation}

In Equation~\ref{eq:beta_aniso}, the \textit{Green-Lagrange} strain tensor $\mathbf{E}=\frac{1}{2}\left[\mathbf{\mathbf{F}^\text{T}\cdot\mathbf{F}}-	\mathbf{I}\right]$, a covariant tensor, is double-contracted with the normal chains direction in the material configuration, $\hat{\mathbf{M}}_\text{ani}$, a tangent vector obtained through pull-back (contravariant) operation of the spatial counterpart, i.e., $\hat{\mathbf{M}}_\text{ani} = \mathbf{F}^\text{-1}\cdot \left[c \, \hat{\mathbf{m}}_\text{ani}\right]$, with $c$ determined so that $\hat{\mathbf{M}}_\text{ani}$ is unit vector. This allows to make the amplification function directly dependent on the strain of the chains in the loading direction. Note that the positive ramp function $\langle \bullet \rangle_+$ prevents amplification of $G_\text{c}$ under compression, but only under tensile loading of the chains.  Furthermore, the factor $\tilde{\beta}_\text{ani}$ and the exponent $\alpha$ are heuristic parameters that establish how the anisotropic critical energy release rate scales with the stretch of the polymer chains. In the present work, we select $\alpha=1$. Eventually, the use of the degradation function in the former equation is necessary to decrease the anisotropic fracture resistance once damage evolves. Otherwise, the very high deformation of damaged elements leads to nonphysical values of $G_\text{c}$, entailing potential damage reversibility. We note that for more complex loading modes Eq.~\ref{eq:beta_aniso} may be modified to not only depend on the principal loading direction $\hat{\mathbf{m}}_\text{ani}$, but on a more representative measure of the deformation state. This may constitute a point for future work.

The contribution $D_\text{crack,visc}$ in Equation~\ref{eq:total_rate_potential_S} allows to model rate-dependent crack growth dissipation, where the scalar $\eta$ denotes a viscosity parameter. The use of a small enough value of $\eta$ even for quasi-static loading is common practice to enhance the numerical robustness of the damage field evolution. 

The first variation of the rate potential functional $\mathrm{F}$ renders the phase-field equation
\begin{align}\label{eq:fieldeq_damage_S}
& \delta_{\dot{d}}\mathrm{F} = \left.\frac{\rm{d}}{\rm{d} \lambda} \mathrm{F}(\dot{d}+\lambda \delta \dot{d})\right\vert_{\lambda=0} = 0
\quad \rightarrow \quad \\
 &  G_\text{c}\left(\mathbf{F},d\right) \frac{3}{8} \left[\frac{1}{l} - 2 l \nabla_0^2 d \right]+g'\left(d\right) \Psi +\eta\dot{d}
  = 0, \nonumber
\end{align}
\noindent with $g'\left(d\right)=2\left[d-1\right]$ the derivative of the degradation function with respect to $d$.

Likewise, the localized quasi-static balance field equation can be obtained from
\begin{equation}\label{eq:fieldeq_mech_S}
\delta_{\dot{\mathbf{u}}} \mathrm{F} = \left.\frac{\rm{d}}{\rm{d} \lambda} \mathrm{F}(\dot{\mathbf{u}}+\lambda  \delta \dot{\mathbf{u}})\right\vert_{\lambda=0} = 0
\quad \rightarrow \quad\\
\nabla_0 \cdot\mathbf{P} = \mathbf{0},
\end{equation}
\noindent where $\mathbf{P}$ refers to the Piola stress tensor.

%\begin{remark}
Note that the rate-dependent potential functional in Equation~\ref{eq:total_rate_potential_S} allows to set the material function $G_\text{c}\left(\mathbf{F},d\right)$ as a function of the primary fields but independent of their rates of change. As a consequence, its variation with respect to $\dot{d}$ and $\dot{\mathbf{F}}$ is zero, i.e., $\partial_{\dot{d}} G_\text{c}\left(\mathbf{F},d\right)=0$ and $ \partial_{\dot{\mathbf{F}}}G_\text{c}\left(\mathbf{F},d\right)=\mathbf{0}$, and Equations \ref{eq:fieldeq_damage} and \ref{eq:fieldeq_mech} contain no additional contributions related to these terms.

Finally, only the constitutive relations for the energy density remain to be defined. The isochoric contribution to the total energy density in Equation~\ref{eq:total_energy_density_S} is defined according to the Yeoh model as
\begin{equation}\label{eq:Psi_isochoric_S}
\Psi_\text{iso}\left(\overline{\mathbf{F}}\right)=
C_{1}\left[\overline{I}_1-3\right]+C_{2}\left[\overline{I}_1-3\right]^2+C_{3}\left[\overline{I}_1-3\right]^3,
\end{equation}
\noindent with the isochoric invariant $\overline{I}_1=\text{tr} \left(\overline{\mathbf{F}}^\text{T}\cdot\overline{\mathbf{F}}\right)$ and $2C_{1}$ the shear modulus. The calibration of the coefficients $C_{1}$, $C_{2}$, and $C_{3}$ parameters is detailed in Methods 3.

For the volumetric contribution, we use a relation directly dependent on the bulk modulus that is adequate to recover the nearly incompressible behavior of elastomers, i.e.,
\begin{equation}
\Psi_{\text{vol}} \left( J \right) =
\frac{\kappa}{2} \left[J-1\right]^2,
\quad 
\textrm{with}
\quad
\kappa=\frac{4C_1\left[1+\nu \right]}{3\left[1-2\nu\right]},
\end{equation}
\noindent for bulk modulus $\kappa$ with Poisson ratio $\nu$ set to 0.49.

Subsequently, the Piola stress tensor can be derived from the energy density as the addition of isochoric and volumetric contributions according to $\mathbf{P}=g\left(d\right)\mathbf{P}_\text{iso}+g\left(d\right)\mathbf{P}_\text{vol}$. The isochoric contribution results
\begin{align}\label{eq:}
\mathbf{P}_\text{iso} = \frac{\partial \Psi_\text{iso} \left(\overline{\mathbf{F}}\right)}{\partial \mathbf{F}} =
J^{-1/3}\mathbb{K}:\frac{\partial \Psi_\text{iso}\left(\overline{\mathbf{F}}\right)}{\partial \overline{\mathbf{F}}}
=  J^{-1/3}\mathbb{K} : 
\left[2C_{1}+4C_{2}\left[\overline{I}_1-3\right]+6C_{3}\left[\overline{I}_1-3\right]^2\right]\overline{\mathbf{F}},
\end{align}
\noindent with the fourth-order mixed-variant projection tensor $\mathbb{K}=\mathbb{I}-\frac{1}{3}\mathbf{F}^{-\text{T}}\otimes\mathbf{F}$, and the volumetric contribution
\begin{align}\label{eq:Pvol_v_S}
\mathbf{P}_\text{vol}
=\frac{\partial \Psi_\text{vol}\left(\mathbf{F}\right)}{\partial \mathbf{F}}=J\frac{\partial \Psi_\text{vol}\left(\mathbf{F}\right)}{\partial J}\mathbf{F}^{-\mathrm{T}}
=\kappa\left[J^2-J\right]\mathbf{F}^{-\text{T}}.
\end{align}

\newpage
\subsection*{\textup{\textbf{Methods 3: Numerical implementation}}}
For the numerical implementation of the field equations, their weak forms need to be derived testing and integrating the field equations. The weak form of the phase-field equation is obtained by multiplying Equation~\ref{eq:fieldeq_damage} times an admissible test function $\delta d$ and integrating by parts
\begin{equation}\label{eq:weakfieldeq_damage_S}
\int_{\Omega_0}
\left[G_\text{c}\left(\mathbf{F},d\right) \frac{3}{8}\left[\frac{\delta d}{l} + 2 l  \nabla_0 d \cdot \nabla_0 \delta d \right]
+ g\left(d\right)' \delta d \Psi
+ \frac{\eta}{\tau}\dot{d}\,\delta d \right] \text{d}V
=0
\end{equation}

Likewise, the weak form of the localized quasi-static force balance is obtained multiplying Equation~\ref{eq:fieldeq_mech} times an admissible test function $\delta \mathbf{u}$ and integrating by parts
\begin{equation}\label{eq:}
\int_{\Omega_0}
 \left[\mathbf{P} 
    :\nabla_0 \delta \mathbf{u}  \right]
    \textrm{d}V= 0
\end{equation}

The weak form of the problem is numerically solved in the open-source finite element environment FEniCS using the NonLinearVariationalProblem library \cite{Logg2012}. The mechanical problem and the phase-field problem are decoupled in a staggered scheme as in \cite{Moreno-Mateos2024a}. Iterations stop when the maximum difference of the fields between two consecutive iterations is smaller than a tolerance, here set to $tol_1=1\cdot10^{-2}$. To decrease the calculation time, we implement an adaptive load stepping that increases and reduces the load step according to the growth of the damage variable. For an increase larger than $tol_2=0.15$, the load increment is reduced by a factor of 1/1.2 up to a minimum value of $\Delta_t=6\cdot10^{-4}$. Otherwise, it is incremented by a factor of 1.2 up to a maximum value of $\Delta_t=1\cdot10^{-2}$. Moreover, Equation~\ref{eq:weakfieldeq_damage_S} is implemented in a staggered manner through an inner loop so that the damage field $d$ in $G_\text{c}\left(\mathbf{F},d\right)$ takes the value of the field in the previous step, i.e., $d_i=d_{i-1}$.

The two-dimensional finite element computations for single-material samples with initial pre-cut of \qty{6}{\milli \meter} were performed on a mesh with \qty{166818} triangular elements. The mesh replicates only the upper half of the samples due to horizontal symmetry. In this regard, we note that sideways cracks do not occur strictly in a symmetric fashion. Nonetheless, we observe that a second sideways crack symmetric to the initial one initiates once the first one stops propagating, and in certain experiments the onset of both cracks occurs simultaneously. With this simplified modeling assumption we preserve the physical behavior while reducing the computational expenses for finer FE meshes.

While 3D simulations are conducted to calibrate the constitutive parameters in Table~\ref{tab:calibration}, 2D simulations are performed for fracture simulations. This allows to reach smaller mesh sizes than with a 3D mesh, while keeping the computational expenses within computation limits and thus the computation times efficiently bounded. The effect of the thickness of the samples on the sideways cracking mechanism is not object of the present study. The reader interested in such effect can consult \cite{Lee2019c}. According to this work, the limited thickness of the samples promotes forward cracking. Although this effect is not captured with our 2D simulations, the error due to this modeling simplification is deemed insignificant and does not hinder the qualitative and quantitative fracture behavior. 

\newpage
\subsection*{\textup{\textbf{Results 1: Results from tensile tests on uncut samples and calibration of the constitutive model}}}
Tensile tests on uncut samples were utilized to calibrate the mechanical material parameters. To enhance the accuracy of this calibration, 3D simulations were conducted to replicate the actual boundary conditions of the specimens. In these simulations, the horizontal displacement at the upper and lower edges was constrained to zero, reflecting the clamping in the experimental setup. The mechanical contribution in the numerical model was calibrated using experimental results from tensile tests on virgin samples, as depicted in Figure~\ref{fig:calib_Yeoh}.A. The parameters for the isochoric contribution, based on the Yeoh model, are listed in Table~\ref{tab:calibration}. The bulk modulus was chosen to be several orders of magnitude higher to guarantee near-incompressibility.
\begin{table}[H]
\caption{Constitutive parameters of the Yeoh model used in the numerical simulations of the elastomer Elastosil P7670 manufactured according to the mixing ratios $\zeta=\{0.5, 0.625, 0.75, 0.875, 1 \}$.}
\label{tab:calibration}
\centering
\definecolor{lightgrey}{RGB}{243, 243, 243}
\definecolor{grey}{RGB}{215, 215, 215}
\begin{tabular}{l|lllll}
\cellcolor{lightgrey} $\zeta$ [-] &\cellcolor{lightgrey} 0.5 &\cellcolor{lightgrey} 0.625 &\cellcolor{lightgrey} 0.75 &\cellcolor{lightgrey} 0.875 &\cellcolor{lightgrey} 1\\ 
\hline 
$C_{1}$~[kPa] & 1.88 & 4.50 & 10.00 & 19.50 & 24.00 \\ 
$C_{2}$~[kPa] & 0.20 & 0.40 & 0.60 & 1.00 & 1.00 \\ 
$C_{3}$~[kPa] & 0 & 0 & 0.02 & 0.02 & 0.08 \\ 
\end{tabular} 
\end{table}

\begin{figure}[h!]
\centering
\includegraphics[width=1\textwidth]{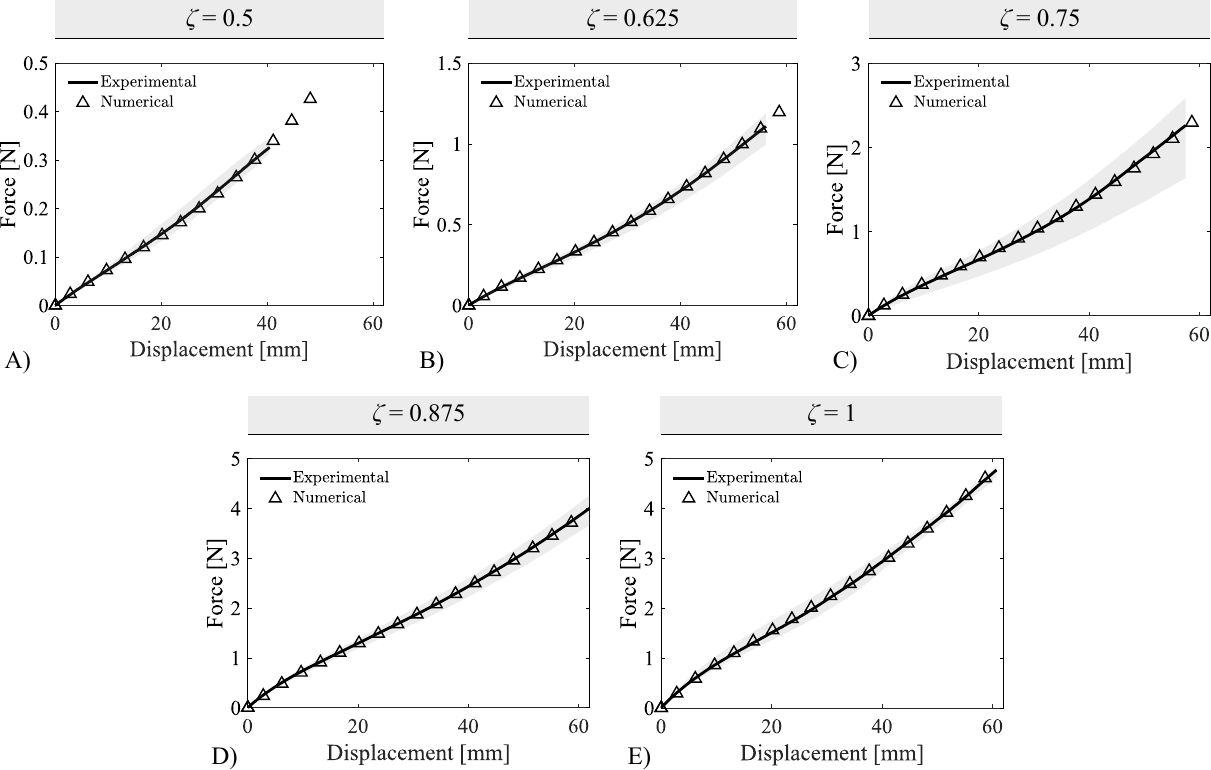}
\caption{\textbf{Calibration of the Yeoh model from quasi-static tensile tests on uncut samples.} A numerical 3D Finite Element model reproduces the experimental setup to bridge the constitutive behavior with the force - displacement measurements obtained through the universal testing machine. The parameters of the Yeoh model, i.e., $C_1$, $C_2$, and $C_3$, are calibrated for samples produced with mixing ratios ($\zeta$) of (A) \qty{0.5}{}, (B) \qty{0.625}{}, (C) \qty{0.75}{}, (D) \qty{0.875}{}, and (E) \qty{1}{}. For each ratio, the experimental mean curve is calculated from three experimental repetitions, whereby a scatter area is included to quantify the experimental variability.
}
\label{fig:calib_Yeoh}
\end{figure}

\newpage
\subsection*{\textup{\textbf{Results 2: Estimation of isotropic critical energy release rate}}}
An estimation of the critical energy release rate ($G_\text{c,iso}$) for forward cracking was performed on samples with low crosslinking ($\zeta=0.5$) according to the definition $G_\text{c,iso}=-\frac{\text{d}U}{t\text{d}c}$, as in \cite{Moreno-Mateos2023,Moreno-Mateos2024a}, with $\text{d}U$ denoting the release of hyperelastic energy for a differential crack extension $\text{d}{U}$ on a sample with a thickness $t$. Here, three estimations are made for finite crack extensions for a $c/w$ from (A) 0.1 to 0.3, from (B) 0.3 to 0.5, and (C) from 0.1 to 0.5. Three similar values are obtained: $G_\text{c,iso}=\qty{51.8e-3}{\newton \per \milli \meter}$, $G_\text{c,iso}=\qty{36.9e-3}{\newton \per \milli \meter}$, and $G_\text{c,iso}=\qty{38.5e-3}{\newton \per \milli \meter}$, respectively. The average value \qty{42.4e-3}{\newton \per \milli \meter} is used. Figure~\ref{fig:calib_Gc} illustrates the results. 
\begin{figure}[h!]
\centering
\includegraphics[width=1\textwidth]{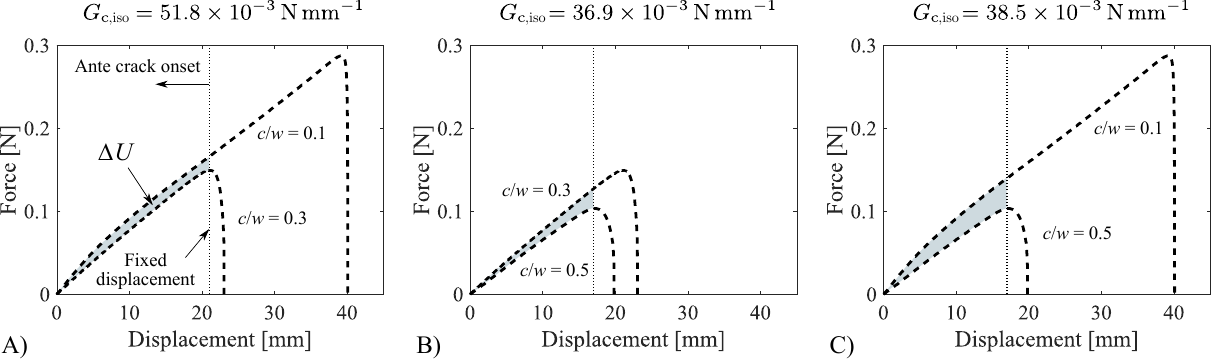}
\caption{\textbf{Estimation of the isotropic critical energy release rate from samples with forward cracking.} Samples prepared for a mixing ratio ($\zeta$) of 0.5 and low degree of crosslink enable estimations of the critical energy release rate as the difference of energy between force displacement curves on samples with pre-cuts of different lengths, defined through the crack width ratio ($c/w$), before crack onset. Three estimations are made for crack extension for a $c/w$ from (A) 0.1 to 0.3, from (B) 0.3 to 0.5, and (C) from 0.1 to 0.5. Three similar values are obtained: $G_\text{c}=\qty{51.8e-3}{\newton \per \milli \meter}$, $G_\text{c}=\qty{36.9e-3}{\newton \per \milli \meter}$, and $G_\text{c}=\qty{38.5e-3}{\newton \per \milli \meter}$, respectively.
}
\label{fig:calib_Gc}
\end{figure}

\newpage
\subsection*{\textup{\textbf{Results 3: Crack paths for samples with different crosslinking degrees and pre-cuts}}}
Figures~\ref{fig:exp_cracks_undef} and \ref{fig:exp_cracks_def}  present the results for crack extension in samples with mixing ratios of $\zeta = \{0.625, 0.75, 0.875, 1\}$, which originate sideways cracking, in the material and spatial configurations, respectively. The data is shown for samples with pre-cuts of lengths $c = \{1.2, 3.6, 6\}~\text{mm}$, corresponding to crack-width ratios of $c/w = \{0.1, 0.3, 0.5\}$.
\begin{figure}[h!]
\centering
\includegraphics[width=0.8\textwidth]{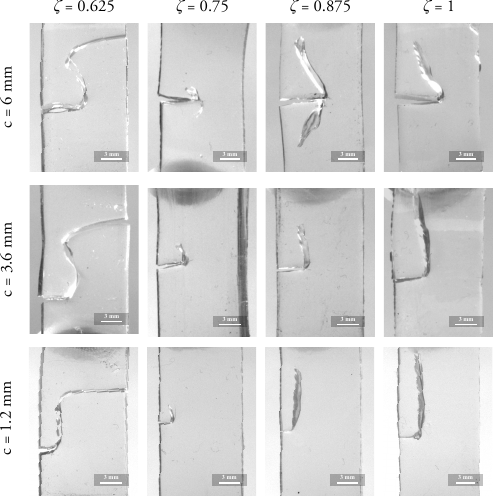}
\caption{\textbf{Results for crack propagation for samples with different mixing ratios and initial pre-cuts in the undeformed configuration.} The images correspond to fully rupture for the low crosslinked sample with $\zeta=0.625$ and to maximum sideways extension for $\zeta=0.75$, \qty{0.875}{}, and \qty{1}{}. Low crosslinked samples for $\zeta=0.5$ are not included since crack propagates forward and the path is the standard horizontal one.
}
\label{fig:exp_cracks_undef}
\end{figure}

\begin{figure}[h!]
\centering
\includegraphics[width=0.6\textwidth]{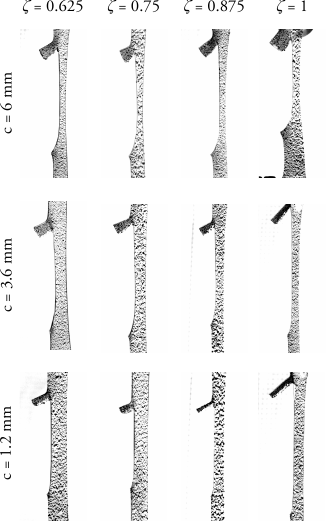}
\caption{\textbf{Results for crack propagation for samples with different mixing ratios before full rupture and initial pre-cuts in the deformed configuration.} The results are presented for the mixing ratios of $\zeta=0.625,0.75,0.875,1$. The results for low crosslinked samples with $\zeta=0.5$ are not included since crack propagates forward and the path is the standard horizontal one.
}
\label{fig:exp_cracks_def}
\end{figure}

\newpage
\subsection*{\textup{\textbf{Results 4: Computationa
l results for anisotropic fracture as a function of $\zeta$ and $\tilde{\beta}_\text{ani}$}}}
The propagation of sideways cracks in the computational model depends on the value of the anisotropy factor $\tilde{\beta}_\text{ani}$. The higher the value, the higher the resistant to forward propagation. To illustrate this dependency, Figure~\ref{fig:comput} shows the crack paths in the material (undeformed) configuration for the mixing ratios $\zeta=0.625,0.75,0.875,1$ and an array of values of  $\tilde{\beta}_\text{ani}$. As the crosslinking degree increases, the amplification of the fracture anisotropy according to Eq.~\ref{eq:gamma_crack_regularisation_S} enhances.
\begin{figure}
\centering
\includegraphics[width=0.7\textwidth]{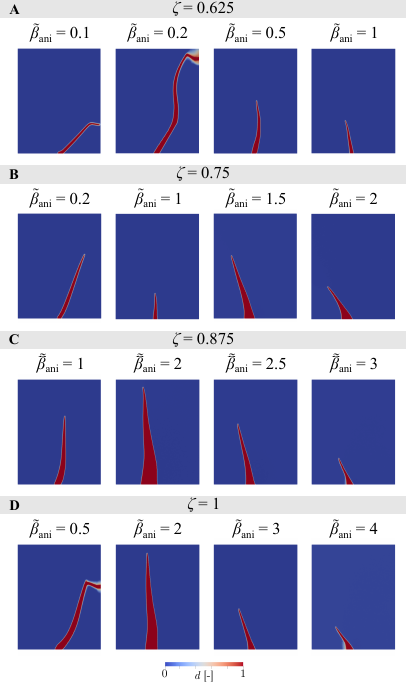}
\caption{\textbf{Results for crack propagation for different parameters in the phase-field model and crack-width ratio $c/w=0.5$.} 
The crack path is shown on the material (undeformed) configuration for sideways cracks produced for mixing ratios ($\zeta$) of (A) 0.625, (B) 0.75, (C) 0.875, and (D) 1, and a variety of values of the anisotropic factor $\tilde{\beta}_\text{ani}$. The increase of the strain induced anisotropy occurs with the increase of this parameter. 
The constitutive behavior of the material varies with the mixing ratios according to the parameters in Table~\ref{tab:calibration}. 
The initial length of the pre-cut, modeled as a discontinuity in the mesh, is \qty{6}{\milli \meter} and the width of the sames ($w$), \qty{12}{\milli \meter}.
}
\label{fig:comput}
\end{figure}

\subsection*{\textup{\textbf{Results 5: Computational results for plane-strain fracture of RVEs with high-crosslinked inclusions embedded in a low-crosslinked soft matrix}}}
Figure~\ref{fig:composite_predictive_bis} showcases the results for the 3D-simulations on Representative Volume Elements (RVE) with inclusions and reinforcements under alternative boundary conditions: $z$-displacement fixed to zero on the $x-y$ plane on the side of the RVE.
\begin{figure}[h!]
\centering
\includegraphics[width=1\textwidth]{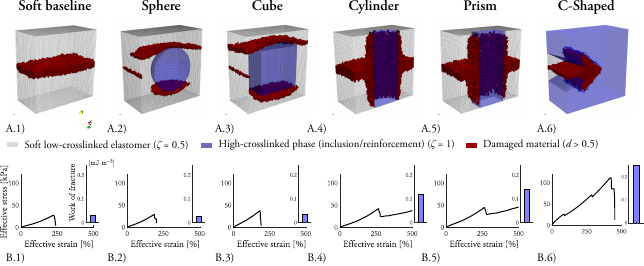}
\caption{\textbf{Results for the fracture behavior of Representative Volume Elements with high-crosslinked inclusions ($\zeta=1$) embedded in a low-crosslinked soft matrix ($\zeta=0.5$).} The \textit{z}- displacement is fixed to zero on the lateral sides ($x-y$ symmetry plane and parallel plane on the side of the RVE) and \textit{x}-displacement is not constrained. Initial damage ($d=1$) is prescribed on one side at the middle of the height. Computational domain in the material (undeformed) configuration where fracture has evolved ($d>0.5$) for a (A.2) spherical inclusion, (A.3) cylindrical fiber, (A.4) cubic inclusion, (A.5) prismatic fiber, and (A.6) C-shaped inclusion. (A.1) A purely low-crosslinked RVE is included as a baseline for comparison to the other cases. The length of the edges of the cubic RVEs is \qty{10}{\milli \meter}, the radius of the centered spherical inclusion \qty{3}{\milli \meter}, \qty{2}{\milli \meter} diameter for the cylindrical fiber, the edge of the cubic inclusion \qty{6}{\milli \meter}, and \qty{4}{\milli \meter} for the prismatic fiber. (B.1-6) Curves with the engineering average stress against the engineering average strain for all cases, including the work of fracture as the area under the curves divided by the total volume of the RVE. For the Cylinder and Prism fiber cases, the work of fracture is calculated up to a displacement of \qty{50}{\milli \meter}.}
\label{fig:composite_predictive_bis}
\end{figure}

\newpage
\subsection*{\textup{\textbf{Video 1: Animation of the deformation and fracture propagation in a C-Shaped RVE obtained with the computational model}}}
}

\newpage
\bibliographystyle{naturemag}%{unsrt}%{ieeetr}
\addcontentsline{toc}{section}{References}
%\bibliography{bibliography}

\end{document}